\documentclass[12pt]{article}
\usepackage[title]{appendix}
\usepackage{mathrsfs} 
\usepackage[bf,small]{caption}
\usepackage[export]{adjustbox}
\usepackage[sort&compress,square,numbers]{natbib}
\usepackage[title]{appendix}
\usepackage[top=2.5cm,left=2.7cm,right=2.7cm,bottom=3.2cm]{geometry}
\usepackage{amsmath,amsthm,amssymb,setspace,enumitem,epsfig,titlesec,verbatim,color,array,multirow,comment,graphicx}
\usepackage{amssymb,amsfonts,amsmath,amsthm}
\usepackage{authblk}
\usepackage{bbm}
\usepackage{fullpage}
\usepackage{graphbox}
\usepackage{graphicx, xcolor}
\usepackage{graphicx}
\usepackage{hyperref}
\usepackage[capitalize]{cleveref}
\usepackage{multicol}
\usepackage{subcaption}
\usepackage{times}
\usepackage{nicefrac}
\usepackage{multirow}

\usepackage{cleveref}
\usepackage[title]{appendix}
\usepackage{mathrsfs} 

\crefname{figure}{Fig}{Figs}
\Crefname{figure}{Fig}{Figs}

\makeatletter
\def\blfootnote{\gdef\@thefnmark{}\@footnotetext}
\makeatother

\newtheorem{theorem}{\bf Theorem}[section]
\newtheorem{lemma}{\bf Lemma}[section]

\newtheorem{corollary}{\bf Corollary}[section]

\newtheorem{definition}{\bf Definition}[section]

\begin{document}

\title{%
  Mixed updating in structured populations%
}
\date{}

\author[a,b,c]{David~A.~Brewster}
\author[a]{Yichen~Huang}
\author[a]{Michael~Mitzenmacher}
\author[c,d]{Martin~A.~Nowak}

\affil[a]{John A. Paulson School of Engineering and Applied Sciences, Harvard University, Boston, MA 02134, USA}
\affil[b]{Department of Molecular and Cellular Biology, Harvard University, Cambridge, MA 02138, USA}
\affil[c]{Department of Organismic and Evolutionary Biology, Harvard University, Cambridge, MA 02138, USA}
\affil[d]{Department of Mathematics, Harvard University, Cambridge, MA 02138, USA}
\maketitle

\blfootnote{Corresponding author: David A. Brewster (\href{mailto:dbrewster@g.harvard.edu}{dbrewster@g.harvard.edu})}

%
%
%
%
%

\begin{abstract}
  Evolutionary graph theory (EGT) studies the effect of population structure on evolutionary dynamics.
  The vertices of the graph represent the $N$ individuals.
  The edges denote interactions for competitive replacement.
  Two standard update rules are death-Birth (dB) and Birth-death (Bd).
  Under dB, an individual is chosen uniformly at random to die, and its neighbors%
  ---the individuals on adjacent vertices of the graph---%
  compete to fill the vacancy proportional to their fitness.
  Under Bd, an individual is chosen for reproduction proportional to fitness,
  and its offspring replaces a randomly chosen neighbor on an adjacent vertex.
  Here we study mixed updating between those two scenarios.
  In each time step, with probability $\delta$ the update is dB and with the remaining probability it is Bd.
  We study fixation probabilities and times as functions of $\delta$ under neutral evolution and constant selection.
  Despite the fact that fixation probabilities and times can be increasing, decreasing, or non-monotonic in $\delta$,
  we prove that nearly all unweighted undirected graphs have short fixation times
  and provide an efficient algorithm to estimate their fixation probabilities.
  Finally, we prove exact formulas for fixation probabilities on cycles, stars, and more complex structures
  and classify their sensitivities to $\delta$.
\end{abstract}

\maketitle

\section*{Author summary}
Evolutionary graph theory models how a population's interaction structure shapes the fate of a newly arising mutant.
Outcomes depend not only on that structure but also on the timing of the birth and death events in each update.
Two standard choices, death-Birth (dB) and Birth-death (Bd), can give strikingly different predictions on the very same structure.
Most previous work studies one rule alone. We mix them, taking each step to be dB with probability $\delta$ and Bd otherwise.
We ask how the fixation probability (the chance a mutant takes over) and the fixation time depend on $\delta$.
Both can behave in surprisingly complicated ways, rising, falling, or reversing direction as $\delta$ changes, yet striking regularities emerge.
For a neutral mutant, the fixation probability averaged over all starting locations is exactly $1/N$ on every structure and for every $\delta$,
and a structure is insensitive to $\delta$ when its individuals all have roughly the same number of neighbors.
When dB and Bd are equally likely ($\delta=1/2$), fixation is fast, for neutral mutants and under constant selection alike.
We also derive exact formulas for cycles and stars, and give an efficient algorithm to estimate fixation probabilities on almost any structure.

\section{Introduction}
In evolutionary graph theory, population structure is described by a weighted or unweighted graph together with an update rule.
Individuals occupy the vertices of the graph, and edges specify potential replacement interactions.
The neighbors of a vertex indicate the possible locations where offspring can be placed.
If the graph is weighted, some replacement events occur with higher probability than others;
a self-loop indicates that an offspring may replace its parent.
Reproduction is typically assumed to be asexual.
A well-mixed population is represented by an unweighted complete graph with self-loops.

Evolutionary graph theory can be used to study
neutral evolution~\cite{diaz_approximating_2014,gao2024speed,brewster2025maintaining},
constant selection~\cite{broom2008analysis,diaz2013fixation,hindersin2015most,adlam2015amplifiers,galanis2017amplifiers,pavlogiannis2017amplification,pavlogiannis2018construction,moller2019exploring,goldberg2019asymptotically,tkadlec2020limits,allen2021fixation,bhaumik2024constant,kuo2024evolutionary,kuo2024theory,brewster2024fixation,kalirad2025ecological,frieze2025moran},
or frequency dependent selection~\cite{ohtsuki2006evolutionary,gomez2007dynamical,broom2010evolutionary,poncela2010cooperation,hadjichrysanthou2011evolutionary,zukewich2013consolidating,wang2013impact,allen2017evolutionary,diaz2021survey,sheng2024strategy,moawad2024evolution}.
The latter is an approach for evolutionary game theory in structured populations.
In this paper, we focus on neutral evolution and constant selection.

Two update rules that have been widely considered are \emph{death-Birth (dB)} updating and \emph{Birth-death (Bd)} updating.
For \emph{death-Birth (dB)} updating, an individual is selected for death uniformly at random among the population.
Then the neighbors compete for the empty site proportional to their fitness and the weight of the link that leads from them to the empty site.
For \emph{Birth-death (Bd)} updating, an individual is selected for reproduction proportional to its fitness, and
then the offspring replaces a neighbor with a probability that is proportional to the weight of the edge that leads to this neighbor (see~\cite{nowak2006evolutionary}).

Past work in evolutionary graph theory has yielded a rich set of results regarding fixation probabilities and fixation times%
~\cite{nowak2003linear,lieberman2005evolutionary,hindersin2015most,adlam2015amplifiers,galanis2017amplifiers,pavlogiannis2017amplification,pavlogiannis2018construction,goldberg2019asymptotically,allen2020transient,sharma2022suppressors,Monk_van_Schaik_2022}.
The fixation probability is the probability that a newly introduced mutant takes over the entire population.
The fixation probability can be calculated for a mutant that starts in a specific place or averaged over all initial placements.
The unconditional fixation time (also known as the absorption time) refers to the average time it takes for a population to reach homogeneity after the introduction of a mutant.
The (conditional) fixation time is the average time it takes for the new mutant to reach fixation. 

One central theme in evolutionary graph theory is that population structure can amplify or suppress selection~\cite{hindersin2015most,adlam2015amplifiers,galanis2017amplifiers,pavlogiannis2017amplification,pavlogiannis2018construction,allen2020transient,sharma2022suppressors,svoboda2024density}.
An amplifier increases the fixation probability of an advantageous mutant and reduces the fixation probability of a deleterious mutant,
whereas a suppressor does the opposite.
Strong amplification comes at the cost of longer fixation times,
so that a structure typically cannot be both a very strong and a very fast amplifier~\cite{tkadlec2019population};
moreover, under death-Birth updating the amplification achievable by any structure is provably bounded~\cite{tkadlec2020limits}.
Beyond amplification, structure also governs how long diversity persists,
with some graphs maintaining diversity for a long time and others eliminating it quickly~\cite{brewster2025maintaining,sharma2025population}.
Crucially, the outcome depends not only on the graph but also on the update rule.
In a well-mixed population, modeled by a complete graph with self-loops, the classic Moran process is recovered regardless of whether Bd or dB updating is used~\cite{moran1958random}.
Yet in structured populations, Bd and dB updating typically lead to different results~\cite{kaveh2015duality,yagoobi2023categorizing,svoboda2024amplifiers}.
This sensitivity to the update rule is precisely what motivates the present work.

In this paper, we study \emph{$\delta$-mixed updating}:
in any one time step, we use \emph{death-Birth (dB)} with probability $\delta$ and we use \emph{Birth-death (Bd)} updating with probability $1-\delta$.
Three earlier lines of work have bridged the two update rules, each approaching the mixture from a different
direction, and together they show that the interpolation is far from a formality.
Zukewich et al.~\cite{zukewich2013consolidating} introduced exactly this mixture for evolutionary games, proving that
on a $k$-regular graph cooperation in the Prisoner's Dilemma is favored precisely when $b/c>k/\delta$ (where $b$ is
the benefit from a cooperator, $c$ is the cost a cooperator pays, and $b>c>0$), so that the dB condition
of~\cite{Ohtsuki_Hauert_Lieberman_Nowak_2006} deforms smoothly across the range of $\delta$, degenerating at
$\delta=0$, where cooperation is never favored.
Tkadlec et al.~\cite{tkadlec2020limits} instead asked what survives of amplification under the same mixture, and
found that the answer changes abruptly at $\delta=0$. Pure Bd updating admits amplifiers, but for every
$\delta>0$ amplification can only be a transient phenomenon with regard to the selection advantage.
Dew and Overton~\cite{dew2025understanding} reach the mixture from an ecological continuous time model, in which an
intrinsic death rate $\Delta$ triggers a death-Birth event while a birth rate $\varphi$ (the average fitness of the
current population) triggers a Birth-death event. After normalizing, their model is $\delta$-mixed updating with
$\delta=\Delta/(\varphi+\Delta)$.
They show that stars on $N$ vertices with self-loops have a critical threshold at $\Delta\sim1/\sqrt{N}$, amplifying
selection under uniform initialization when $\delta\ll1/\sqrt{N}$ and suppressing it when $\delta\gg1/\sqrt{N}$.
Common to all three, however, is that each confines itself to a narrow class of structures (regular graphs, or
stars), and none of them analyzes fixation times.
Our aim is to treat $\delta$ as a parameter on arbitrary structures. We study fixation probability and fixation time
on unweighted undirected graphs as a function of $\delta$, for both neutral evolution and constant selection.

\section{Methods}
\label{sec:methods}
Here we give the formal definitions of the mixed $\delta$-updating model and summarize the analytical tools used to establish our results.

  \subsection{Setup}
  Let $G=(V, E)$ be a connected undirected graph on $N:=|V|$ vertices.
  The number of neighbors of a vertex in a graph is called its \emph{degree}.
  The degree of vertex $u\in V$ is denoted $\deg (u)$.
  On each vertex of $G$ lies an individual that is either a wild type or a mutant type.
  Mutants have relative fitness $r > 0$ and wild-types have fitness $1$.
  When $r = 1$ we are in neutral evolution, when $r > 1$ mutants have a selective advantage (constant selection),
  and when $r < 1$ they are deleterious.
  Our results concern neutral evolution and non-deleterious mutants ($r\geq 1$), with one deliberate exception, namely that
  the exact fixation probability formulas on cycles and stars (\Cref{thm:cycles} and \Cref{thm:stars-selection})
  are stated for all $r>0$, since they remain valid there,
  and \Cref{fig:cycle} and \Cref{fig:star-selection} plot that whole range.

  \subsection{Dynamics}
  Initially, only a subset $S_0\subseteq V$ of the graph is occupied by mutants.
  Then at discrete time steps $t=1, 2, \ldots$, a coupled birth \& death occurs.
  We call subset of mutants $S_t\subseteq V$ the \emph{configuration} of the process after $t$ steps.
  The ordering of these coupled events matters.
  There are two possible orderings we consider.

  \medskip
  \noindent\textbf{death-Birth (dB).}
  A \emph{death-Birth (dB)} step proceeds as follows.
  We first select an individual $v\in V$ for death uniformly at random.
  Then we select a neighboring vertex $u\in V$ of $v$ to give birth proportional to fitness.
  Then the individual at location $u$ places a copy of itself at the recent vacancy of location $v$.
  
  \medskip
  \noindent\textbf{Birth-death (Bd).}
  A \emph{Birth-death (Bd)} step proceeds as follows.
  We first select an individual $u\in V$ for birth proportional to fitness.
  Then we select a neighboring vertex $v\in V$ of $u$ to die uniformly at random.
  The individual at location $u$ then places a copy of itself (and thus of its type) at the recent vacancy of location
  $v$.

  \medskip
  \noindent We will refer to dB and Bd as \emph{pure} updating rules since the updating rule is fixed for each step.
  Next, we define an interpolation of the pure updating rules.

\begin{figure}
    \centering
    \includegraphics[width=\textwidth]{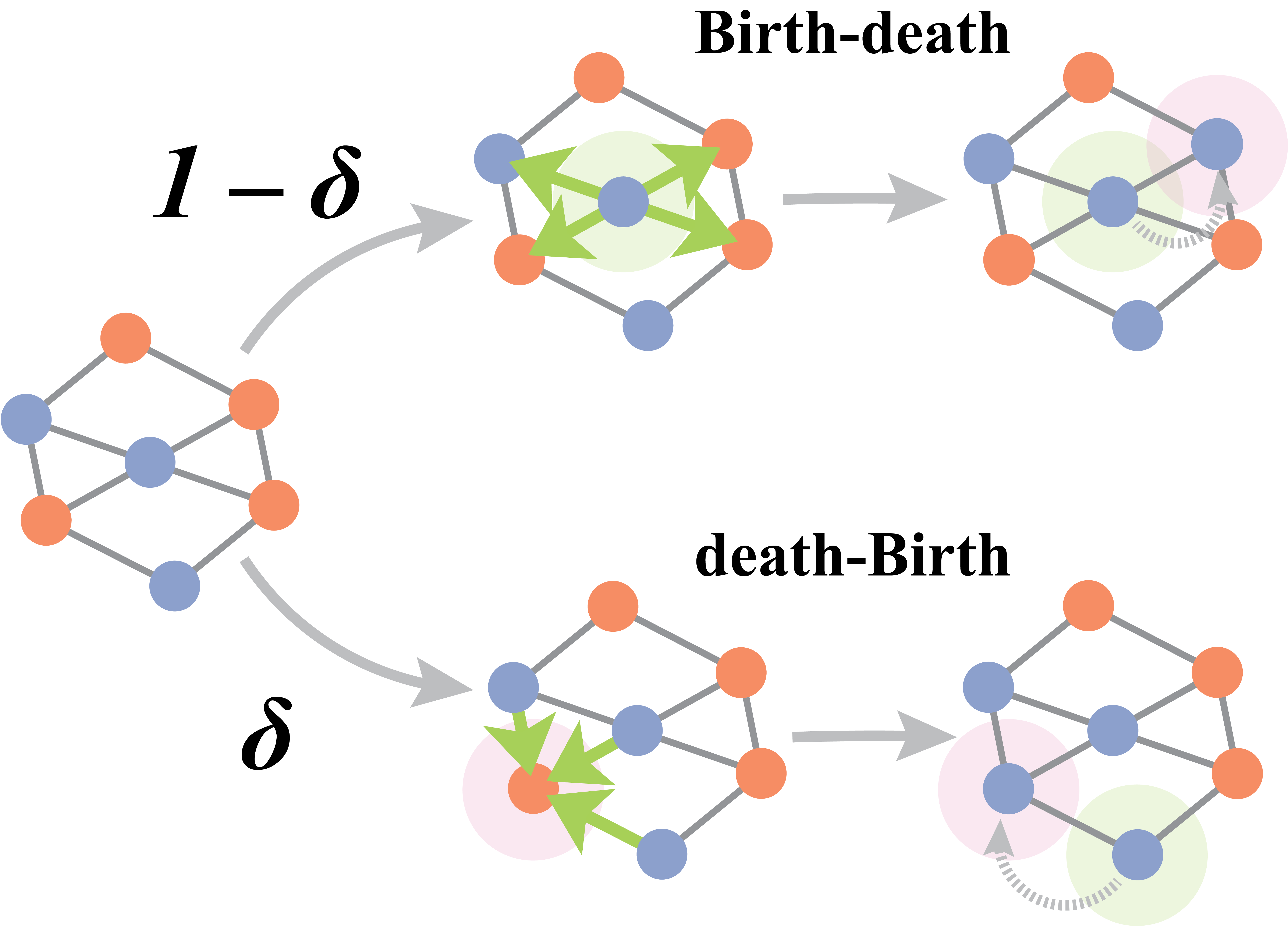}
    \caption{
      Mixed $\delta$-updating on a graph.
      The population consists of $N$ individuals. In this illustration $N=7$ and four of them are mutants.
      The orange vertices are mutants and the blue vertices are wild-type individuals.
      At each time step, a dB step occurs with probability $\delta$,
      and with the remaining probability a Bd step occurs.
      The thick arrows on the edges of the graph are oriented such that the arrowheads indicate the possible location(s)
      of death.
      The arrow tails indicate the possible location(s) of birth for each time step.
      The two update rules generally act differently:
      starting from the same configuration, a dB step and a Bd step can lead to different resulting configurations, as illustrated here.%
    }
    \label{fig:model}
  \end{figure}

  \medskip
  \noindent\textbf{Mixed $\delta$-updating.}
  Let $\delta \in [0,1]$ be the probability a given time step will be a dB step.
  With the remaining probability, the step will be a Bd step.
  We only consider the case when the value of $\delta$ is constant throughout the entire process.
  Note that we recover pure dB updating when $\delta=1$
  and recover pure Bd updating when $\delta=0$.
  When $\delta=1/2$, we refer to the process as \emph{unbiased}.
  See Fig~\ref{fig:model} for an illustration of each step of the process.

  \subsection{Fixation probability and time}
  Eventually, the population becomes entirely mutant (we call this \emph{fixation}) or entirely wild-type (we call this \emph{extinction}).
  In either case, we say that \emph{absorption} has occurred.
  We denote the probability of fixation given mutant fitness $r>0$ and initial set $S_0\subseteq V$
  as $\mathsf{fp}_r^\delta(G, S_0)$.
  We also denote the uniform initialization fixation probability as
  $\mathsf{fp}_r^\delta(G) := \frac1N\sum_{u\in V}\mathsf{fp}_r^\delta(G, \{u\})$.
  When dealing with times, we focus solely on the expected time rather than the entire distribution.
  For an initial mutant set $S_0\subseteq V$,
  we denote the (expected) time until fixation, extinction, and absorption as
  $\mathsf{FT}_r^\delta(G, S_0)$, $\mathsf{ET}_r^\delta(G, S_0)$, and $\mathsf{AT}_r^\delta(G, S_0)$, respectively.
  Since we are interested in upper bounds for times,
  we drop the mutant initial set from the notation for time to denote the maximum time over all possible initial mutant sets.
  Thus, for example, $\mathsf{FT}_r^\delta(G) := \max_{S_0\subseteq V} \mathsf{FT}_r^\delta(G, S_0)$.
  In the notations for probabilities and times, we use $\delta=\textrm{dB}$ to mean $\delta=1$ and $\delta=\textrm{Bd}$ to mean $\delta=0$.
  When the fitness is neutral ($r=1$), we write $\mathsf{fp}_{r=1}^\delta$, $\mathsf{FT}_{r=1}^\delta$, etc.

  \subsection{Analytical techniques}
  Our proofs rely on a potential function approach using martingales.
  For a suitable potential function $\phi$ defined on the mutant configurations $S\subseteq V$,
  we track how $\phi$ changes in expectation under a single mixed $\delta$-updating step and apply optional stopping theorem arguments to compute fixation probabilities and bound absorption and fixation times.
  The formal tools together with all detailed proofs are described in the appendix.

\section{Results}
Throughout this section we use the mixed $\delta$-updating model and the notation for fixation probabilities and times formally defined in the Methods (\Cref{sec:methods}).
Detailed proofs of all results are in the appendix.
\subsection{Neutral evolution}

Suppose that $G=(V,E)$ is a graph and the initial set of mutants resides at $S_0\subseteq V$.
For the pure Bd process in neutral evolution,
it is known that $\mathsf{fp}_{r=1}^{\mathrm{Bd}}(G, S_0) \propto \sum_{u\in S_0} 1/\deg(u)$~\cite{broom2010two};
for the pure dB process,
it is known that $\mathsf{fp}_{r=1}^{\mathrm{dB}}(G, S_0) \propto \sum_{u\in S_0} \deg(u)$~\cite{svoboda2024amplifiers}.
Due to the additivity of the fixation probability in the pure processes, $\mathsf{fp}_{r=1}^{\mathrm{Bd}}(G)=\mathsf{fp}_{r=1}^{\mathrm{dB}}(G)=1/N$.

\begin{figure}
  \includegraphics[width=\textwidth]{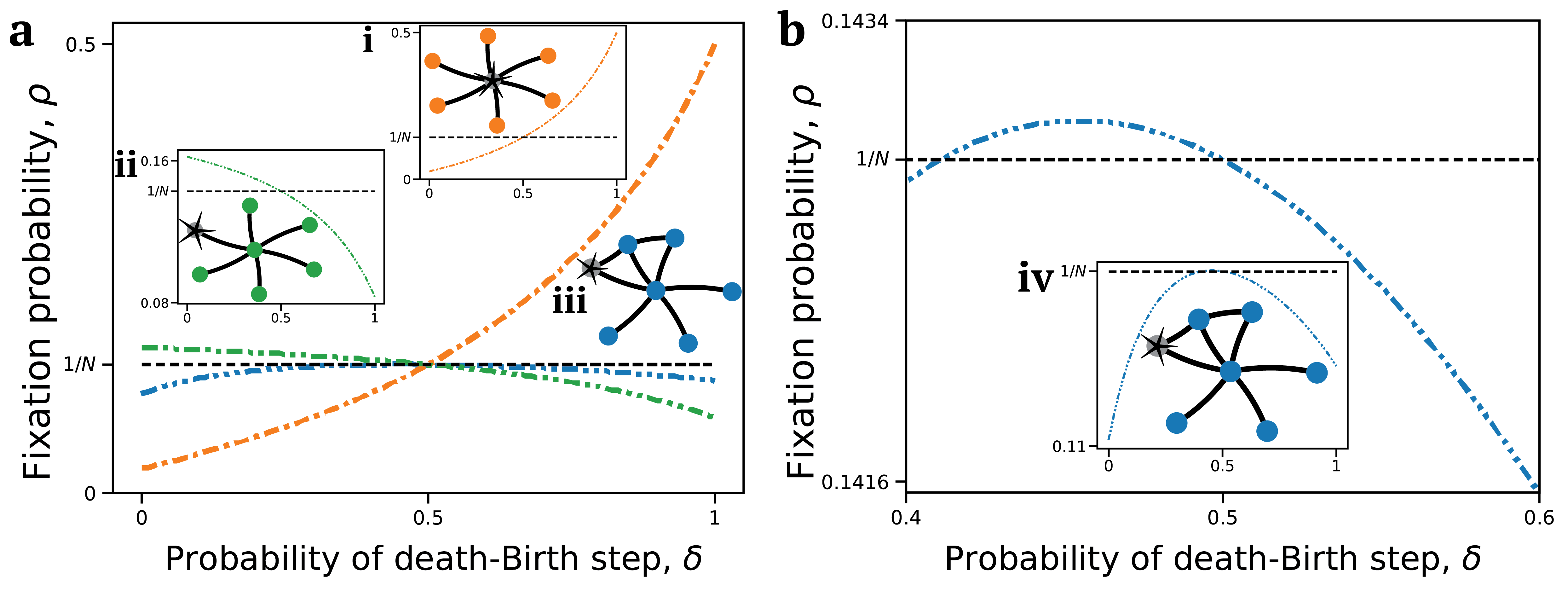}
  \centering
  \caption{%
    Fixation probabilities, $\rho$, of various graphs under mixed $\delta$-updating for $N=7$ in neutral evolution.
    In both panels, the dashed line at $1/N$ is the fixation probability of the Moran process (well-mixed) in neutral evolution.
    For each graph depicted, the small star symbol on the differently shaded vertex indicates the initial mutant location.
    \textbf{a,} \textbf{i)}
    A star graph with the initial mutant location at the center.
    Relative to $\rho$ under the unbiased process, Bd-biased processes have a lower $\rho$ but dB-biased processes have a higher $\rho$.
    \textbf{a,} \textbf{ii)}
    A star graph with the initial mutant location at a leaf.
    Relative to $\rho$ under the unbiased process, Bd-biased processes have a higher $\rho$ but dB-biased processes have a lower $\rho$.
    \textbf{a,} \textbf{iii)}
    A graph composed of a star and two additional edges.
    Relative to $\rho$ under the unbiased process, both pure Bd and pure dB processes have a lower $\rho$.
    However, panel \textbf{b, iv)} shows that the graph from \textbf{a, iii)} has higher $\rho$ than the unbiased process when $\delta$
    is slightly smaller than $1/2$.
    This example shows that $\rho$ can be non-monotonic in $\delta$.
    Further, it is not possible that both pure Bd and pure dB have higher $\rho$ than the unbiased process
    (see~\Cref{thm:both-pure-larger-than-unbiased-not-possible} in \Cref{sec:not-possible}).%
  }
  \label{fig:weird-fp}
\end{figure}

First, we prove that under general $\delta$, the fixation
probability is similarly additive in the initial mutant set, as it is in the pure Bd and dB processes;
that is, $\mathsf{fp}_{r=1}^{\delta}(G, S_0) = \sum_{u\in S_0}\mathsf{fp}_{r=1}^{\delta}(G, \{u\})$
(see~\Cref{claim:fp-additive} in \Cref{sec:additivity}).
Since $\mathsf{fp}_{r=1}^{\delta}(G)=(1/N)\cdot\sum_{u\in V} \mathsf{fp}_{r=1}^{\delta}(G, \{u\})$,
additivity easily yields the following result.
\begin{theorem}[Uniform initialization]
  \label{thm:uniform-init}
  Averaging uniformly over all initial conditions,
  the fixation probability of a neutral mutant is $1/N$ on any graph
  for any value of $\delta$.
\end{theorem}
\noindent In our notation, \Cref{thm:uniform-init} means that $\mathsf{fp}_{r=1}^{\delta}(G)=1/N$ for any graph $G$ and any value of $\delta$.
It follows immediately from the additivity result~\Cref{claim:fp-additive}, proved in \Cref{sec:additivity}.
Thus mixed $\delta$-updating under uniform initialization is well-behaved.
Let $u\in V$.
Combining the two facts above, we show that if one of $\mathsf{fp}_{r=1}^{\mathrm{Bd}}(G, \{u\})$ or $\mathsf{fp}_{r=1}^{\mathrm{dB}}(G, \{u\})$ is larger than $1/N$,
then the other is at most $1/N$ (see~\Cref{thm:both-pure-larger-than-unbiased-not-possible} in \Cref{sec:not-possible}).
Surprisingly, we prove that no matter the fixation probabilities of the pure processes with initial mutant location $u$,
the unbiased process always has fixation probability $1/N$ when the initial mutant location is $u$.
This means that even if both pure processes have fixation probability less than $1/N$ for initial mutant location $u$,
the unbiased process has a higher fixation probability than both the pure processes for initial mutant location $u$.
\begin{theorem}[Arbitrary initialization]
  \label{thm:neutral-evolution}
  If $\delta=1/2$ (which is the unbiased case)
  then the fixation probability of a neutral mutant starting in an arbitrary location is $1/N$ on any graph.
\end{theorem}
\noindent In our notation, \Cref{thm:neutral-evolution} states that for $G=(V,E)$ and $u\in V$ we have $\mathsf{fp}_{r=1}^{\delta=1/2}(G, \{u\}) = 1/N$.
The proof is in~\Cref{sec:proof-neutral-evolution}.
As demonstrated by Fig~\ref{fig:weird-fp},
the quantity $\mathsf{fp}_{r=1}^{\delta}(G, \{u\})$ as a function of $\delta$ can be increasing, decreasing, concave, convex, or non-monotone.

\begin{figure}
  \centering
  \includegraphics[width=\textwidth]{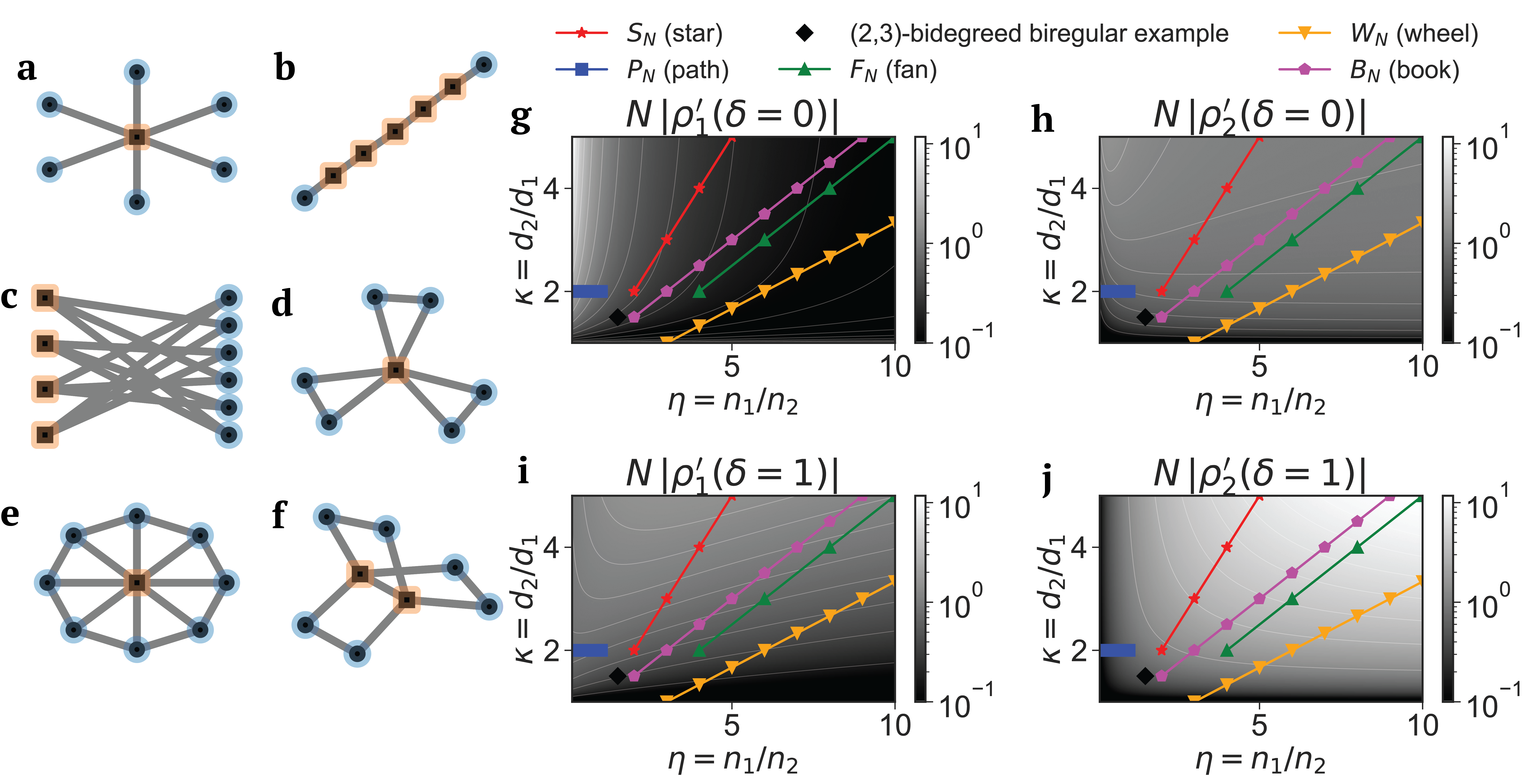}
  \caption{%
    Examples of $(d_1,d_2)$-bidegreed graph families (\textbf{a-f})
    and the size normalized sensitivity of their neutral fixation probabilities to $\delta$,
    shown separately for each initial mutant location and evaluated at $\delta=0,1$ (\textbf{g-j}).
    Vertices with degree $d_1$ are represented as circles whereas
    vertices with degree $d_2\geq d_1$ are represented as squares.
    For the examples depicted, $d_2$ is strictly greater than $d_1$, but it is also possible for $d_2=d_1$ per~\Cref{def:bidegreed}.
    Let $n_1$ denote the number of vertices with degree $d_1$
    and $n_2$ denote the number of vertices with degree $d_2$,
    and set $\eta=n_1/n_2$ and $\kappa=d_2/d_1$.
    The value $\rho_i(\delta):=\mathsf{fp}_{r=1}^\delta(G, \{u\})$ is the fixation probability of a neutral mutant starting at a vertex $u$ of degree $d_i$,
    its derivative with respect to $\delta$ is denoted $\rho'_i(\delta)$, and the sensitivity is $s_i(\delta):=|\rho'_i(\delta)|$.
    Panels \textbf{g--j} display the size normalized sensitivity $N\,s_i(\delta)=N\,|\rho'_i(\delta)|$ evaluated at $\delta=0$ and $\delta=1$.
    Normalizing by $N$ makes the sensitivity a function of $(\eta,\kappa)$ alone,
    since by~\Cref{thm:sensitivity} the sensitivity $s_i(\delta)$ scales as $1/N$ at fixed $(\eta,\kappa)$.
    A graph of size $N$ sitting at a plotted point therefore has sensitivity equal to the plotted value divided by $N$.
    \textbf{a,}
    a star graph, $S_N$. The $n_2=1$ center vertex has degree $d_2=6$
    while the $n_1=6$ periphery vertices have degree $d_1=1$
    for a total of $N=n_1+n_2=6+1=7$ vertices.
    \textbf{b,}
    a path graph, $P_N$. The intermediate vertices have degree $2$ while the endpoint vertices have degree $1$.
    \textbf{c,}
    a bipartite biregular graph; the vertices on the left side have degree $d_2=3$ while the vertices on the right side have degree $d_1=2$, for a total of $N=10$ vertices.
    \textbf{d,}
    a fan graph $F_N$. The center vertex has degree $2b=6$, while each of the $2$ vertices on each of the $b=3$ blades has degree $2$, for a total of $N=2b+1=7$ vertices.
    \textbf{e,}
    a wheel graph $W_N$. The center vertex has degree $n=8$ while the $n$ vertices on the spoke ends have degree $3$, for a total of $N=n+1=9$ vertices.
    \textbf{f,}
    a book graph $B_N$. Each of the $p=3$ pages has $2$ page corners with degree $2$, and
    the $2$ seam endpoints have degree $p+1$, for a total of $N=2p+2=8$ vertices.
    \textbf{g,}
    A contour plot in the $(\eta,\kappa)$ space, where both parameters are dimensionless.
    The lines with shapes denote the values of $\eta$ and $\kappa$ for the various bidegreed graphs in \textbf{a-f}.
    Note that the path graph has $n_1=2$ and $n_2=N-2$, so $n_1\leq n_2$ for $N\geq 4$. Thus as $N$ increases, $\eta$ decreases.
    The intensity of the colorbar denotes $N\,|\rho'_1(\delta=0)|$,
    \textbf{h,}
    $N\,|\rho'_2(\delta=0)|$,
    \textbf{i,}
    $N\,|\rho'_1(\delta=1)|$, and
    \textbf{j,}
    $N\,|\rho'_2(\delta=1)|$.
    Lighter colors indicate a larger sensitivity.
  }
  \label{fig:bidegreed}
\end{figure}

\begin{figure}
  \centering
  \includegraphics[width=\textwidth]{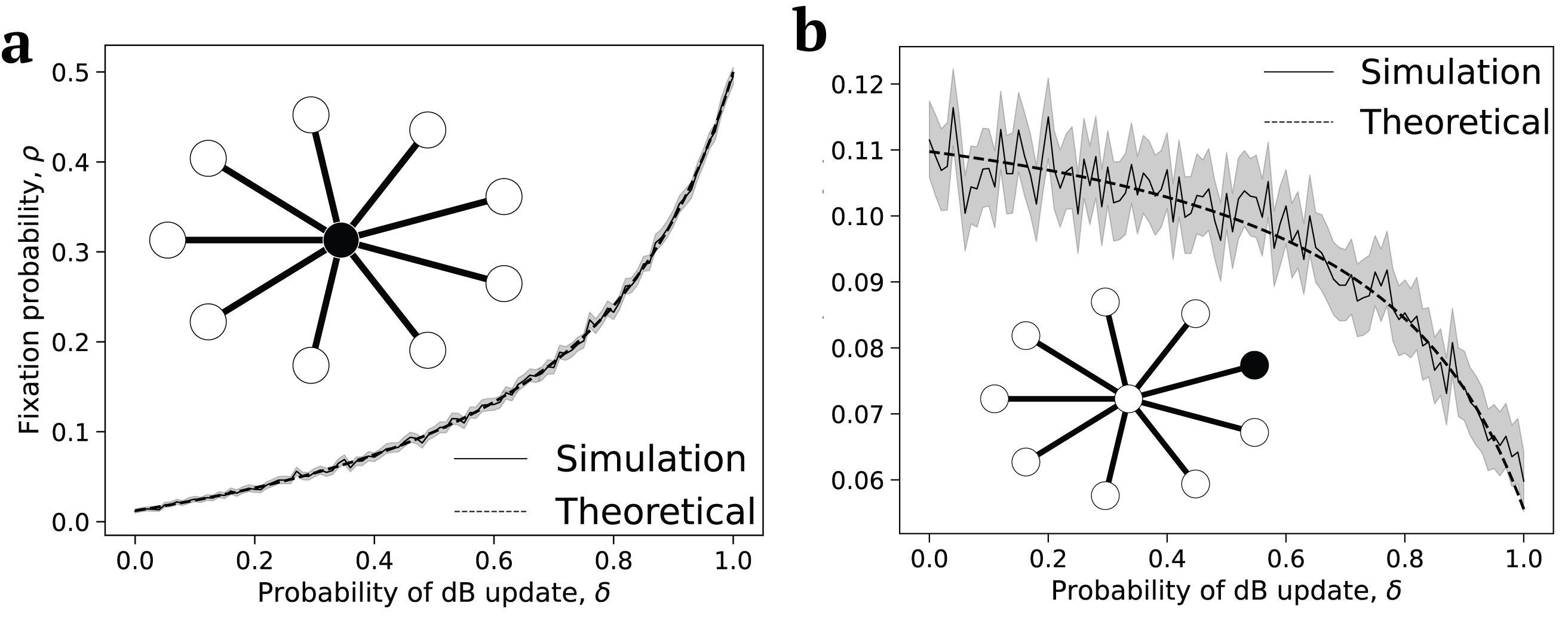}
  \caption{%
    Fixation probabilities on a star graph with $N=10$ vertices in neutral evolution, plotted as a function of $\delta$.
    In each panel, the dashed line is the theoretical fixation probability from~\Cref{thm:stars},
    and the solid line with the surrounding translucent region is the average and its $95\%$ confidence interval from $10^4$ simulations of the mixed $\delta$-updating process.
    \textbf{a,}
    Fixation probabilities as a function of $\delta$ when starting with a mutant in the center.
    \textbf{b,}
    Same as a) except a mutant is initially on a leaf.
  }
  \label{fig:star}
\end{figure}

\noindent Despite this potentially strange behavior, we prove that the fixation probabilities on regular graphs are not affected by $\delta$.
\begin{theorem}[Fixation probability of regular graphs]
  \label{thm:regular-graphs}
  The fixation probability of a neutral mutant starting in an arbitrary location on a regular graph is $1/N$ for any value of $\delta$.
\end{theorem}
\noindent This is a special case of~\Cref{thm:bidegreed} (proved in~\Cref{sec:bidegreed-main}).
In fact, we prove something stronger. There is a broad class of graphs whose
fixation probabilities are monotone in $\delta$.
Such graphs, called \emph{bidegreed}, are those in which every vertex has one of only two possible degrees.
\begin{definition}[Bidegreed graph]
  \label{def:bidegreed}
  A bidegreed graph is an undirected graph with the following property:
  for each vertex the number of neighbors is either $d_1$ or $d_2$.
\end{definition}
\noindent If $d_1=d_2$ then the graph is regular, and
in that degenerate case we adopt the convention $n_1=N$ and $n_2=0$, so that $n_1+n_2=N$ throughout.
\begin{theorem}[Bidegreed graphs]
  \label{thm:bidegreed-behavior}
  On a bidegreed graph with $d_1<d_2$,
  the fixation probability of a neutral mutant starting on a vertex with degree $d_1$ is a strictly decreasing function of $\delta$.
  The fixation probability of a neutral mutant starting on a vertex with degree $d_2$ is a strictly increasing function of $\delta$. 
\end{theorem}
\noindent The proof is in~\Cref{sec:sensitivity}, where items 1--3 below are established as~\Cref{cor:sensitivity-endpoints}.
\noindent Let $n_i$ denote the number of vertices in a bidegreed graph $G$ with degree $d_i$ for $i=1,2$.
We denote $\rho_i(\delta)$ as $\mathsf{fp}_{r=1}^\delta(G, \{u\})$
where $u$ is a vertex with degree $d_i$.
The sensitivity of a vertex with degree $d_i$ is $s_i(\delta):=|\rho'_i(\delta)|$.
In addition to \Cref{thm:bidegreed-behavior}, we prove that for $i=1,2$,
\begin{enumerate}
  \item if $d_1<d_2$ and $n_1>n_2$, $s_i$ has its maximum at $\delta=1$ and minimum at $\delta=0$;
  \item if $d_1<d_2$ and $n_1<n_2$, $s_i$ has its maximum at $\delta=0$ and minimum at $\delta=1$;
  \item if $n_1=n_2$ or $d_1=d_2$, then $s_i$ is constant with respect to $\delta$.
\end{enumerate}
\noindent See Fig~\ref{fig:bidegreed} for examples of bidegreed graphs and their behaviors with respect to $\delta$.
A star graph is the canonical bidegreed graph with extreme degrees (see~\Cref{fig:bidegreed}\textbf{a}).
Key to our proof of \Cref{thm:bidegreed-behavior} is an explicit formula for the fixation probability on a bidegreed graph for any $\delta$.
\begin{theorem}[Bidegreed fixation probabilities]
  \label{thm:bidegreed}
  Let $G$ be a bidegreed graph with degrees $d_1\leq d_2$, having $n_1$ vertices of degree $d_1$ and $n_2$ vertices of degree $d_2$.
  Define
  \begin{equation}
    B_2(\delta) := \frac{(1-\delta)d_1+\delta d_2}{(1-\delta)d_2+\delta d_1}
    \qquad\text{and}\qquad
    S(\delta) := n_1 + n_2\, B_2(\delta).
  \end{equation}
  Then for all $\delta\in[0,1]$, the fixation probability of a neutral mutant starting on a vertex $u$ is
  \begin{equation}
    \mathsf{fp}_{r=1}^\delta(G,\{u\}) =
    \begin{cases}
      \dfrac{1}{S(\delta)} & \text{if } \deg(u)=d_1,\\[10pt]
      \dfrac{B_2(\delta)}{S(\delta)} & \text{if } \deg(u)=d_2.
    \end{cases}
  \end{equation}
  In particular, the ratio of the fixation probability of a degree $d_2$ vertex to that of a degree $d_1$ vertex equals
  $B_2(\delta)=\frac{(1-\delta)d_1+\delta d_2}{(1-\delta)d_2+\delta d_1}$.
\end{theorem}
\noindent The proof is in~\Cref{sec:bidegreed-main}.
A star graph $S_N$ is a canonical example of a bidegreed graph.
With center $c$ and $n=N-1$ leaves $\ell_1,\ldots,\ell_n$, \Cref{thm:bidegreed} gives a formula for the fixation probability for any $\delta$.
We can also state it explicitly.
\begin{theorem}[Fixation probability on stars, neutral evolution]
  \label{thm:stars}
  For a star $S_N$ with center $c$ and $n=N-1$ leaves, for each $i\in\{1,\ldots,n\}$ and all $\delta\in[0,1]$,
  \begin{equation}
    \mathsf{fp}_{r=1}^{\delta}(S_N, \{\ell_i\})=\frac{1+(1-\delta)(N-2)}{(1-\delta)(N-2)^2+2(N-1)}
    \quad
    \text{and}
    \quad
    \mathsf{fp}_{r=1}^{\delta}(S_N, \{c\})=\frac{1+\delta(N-2)}{(1-\delta)(N-2)^2+2(N-1)}.
  \end{equation}
\end{theorem}
\noindent The proof is in~\Cref{sec:star-neutral}.
See Fig~\ref{fig:star} for these two fixation probabilities as a function of $\delta$, together with simulations.

\begin{figure}
  \includegraphics[width=\textwidth]{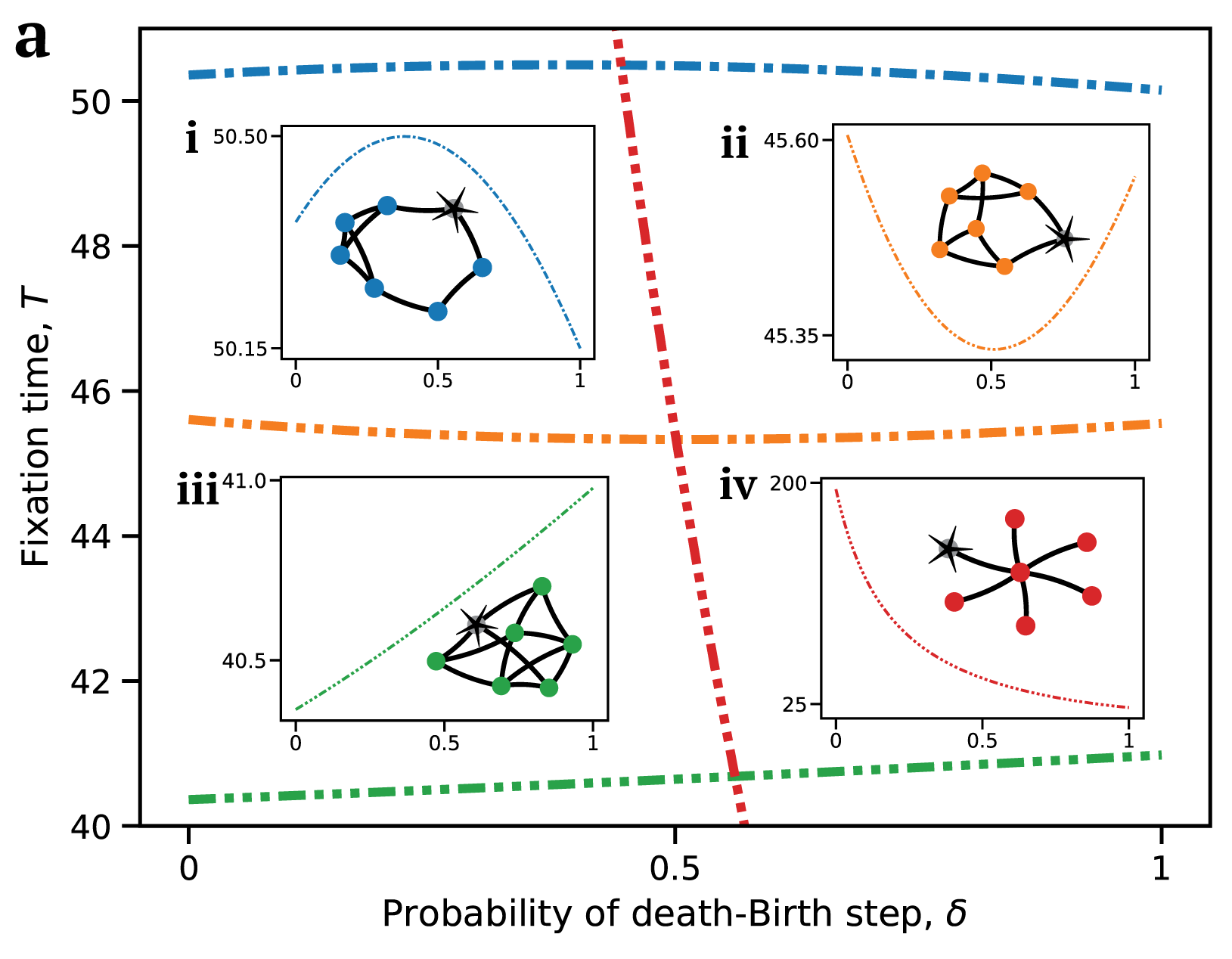} 
  \centering
  \caption{%
    \textbf{a,} Fixation times, $T$, of various graphs under mixed $\delta$-updating for $N=7$ in neutral evolution.
    For each graph depicted, the small star symbol on the differently shaded vertex indicates the initial mutant location.
    \textbf{i)}
    The fixation time increases and then decreases relative to $\delta$.
    \textbf{ii)}
    The fixation time decreases and then increases relative to $\delta$.
    \textbf{iii)}
    The fixation time always increases relative to $\delta$.
    \textbf{iv)}
    The fixation time always decreases relative to $\delta$.
  }
  \label{fig:weird-ft}
\end{figure}

\noindent Similar to fixation probabilities,
fixation times in the mixed $\delta$-updating process can exhibit a large range of behaviors relative to $\delta$ (see Fig~\ref{fig:weird-ft}).
For any sequence of graphs $(G_N)_{N\geq 1}$, one of each size $N$,
it is known $\mathsf{AT}_{r=1}^{\textrm{Bd}}(G_N)$ is $O(N^6)$~\cite{diaz_approximating_2014}.
For pure dB, it is known $\mathsf{AT}_{r=1}^{\textrm{dB}}(G_N)$ is $O(N^5)$~\cite{cooper2016linear}.
Relating fixation times to absorption times,
the law of total expectation on the absorption time gives the bound
$\mathsf{FT}_{r=1}^{\delta}(G,S_0)\leq \mathsf{AT}_{r=1}^{\delta}(G,S_0) / \mathsf{fp}_{r=1}^\delta(G,S_0)$
for any $S_0\subseteq V$ and $\delta\in[0,1]$.
We prove that the absorption and fixation times are well-behaved in the unbiased process.
\begin{theorem}[Fixation time]
  \label{thm:fixation-time-unbiased-neutral}
  If $\delta=1/2$ (which is the unbiased case),
  then the fixation time of a neutral mutant starting in an arbitrary location on any graph of size $N$ is at most $N^5/4$. 
\end{theorem}
\noindent The proof is in~\Cref{sec:fix-time-unbiased}.
\noindent In our notation, \Cref{thm:fixation-time-unbiased-neutral} states that
for all graphs $G=(V,E)$ on $N$ vertices and any initial mutant location $u\in V$,
we have $\mathsf{FT}_{r=1}^{\delta=1/2}(G, \{u\}) \leq N^5/4$.
In the case of $(d_1,d_2)$-bidegreed graphs, we are able to say more about their behavior relative to $\delta$.
In particular, fixation times on bidegreed graphs are at most a polynomial in $N$ and the degree ratio $d_2/d_1$,
uniformly in $\delta$.
\begin{theorem}[Fixation time on bidegreed graphs]
  \label{thm:fixation-time-bidegreed}
  On a bidegreed graph with $d_1\leq d_2$, and for every $\delta\in[0,1]$,
  the fixation time of a neutral mutant starting on a vertex with degree $d_1$
  is at most $N^5\cdot (d_2/d_1)^5 / 4$.
  The fixation time of a neutral mutant starting on a vertex with degree $d_2$
  is at most $N^5\cdot (d_2/d_1)^6 / 4$.
\end{theorem}
\noindent The proof is in~\Cref{sec:fix-time-bidegreed}.
For both \Cref{thm:fixation-time-unbiased-neutral} and \Cref{thm:fixation-time-bidegreed},
those proofs also give the absorption times starting from an arbitrary location, and more precise bounds.

\subsection{Constant selection}

Not only do we obtain results for neutral evolution, but also for constant selection.
Suppose mutants have relative fitness $r > 1$ and wild-types have fitness $1$.
We examine two regimes: weak selection ($r > 1$ on any graph) and strong selection ($r > (D/d)^2$
where $D$ is the maximum degree and $d$ is the minimum degree of the graph).
The two exact formulas below, on cycles and stars, are the exception noted in the Methods, in that
they are stated for all $r>0$, and so also cover deleterious mutants.
Despite the fact that fixation probabilities and times can be increasing, decreasing,
or non-monotonic in $\delta$, we prove that nearly all graphs have short fixation times
and provide an efficient algorithm to estimate fixation probabilities.

\subsubsection{Weak selection}

Here $(G_N)_{N\geq 1}$ denotes an arbitrary sequence of undirected unweighted graphs, one graph $G_N$ of each size $N$ (not a specific named family);
the asymptotic statements below hold uniformly over all such sequences.
It is known that for the pure Bd process with $r\neq 1$, $\mathsf{AT}_r^{\textrm{Bd}}(G_N)$ is $o(N^{3+\varepsilon})$
for all $\varepsilon>0$~\cite{diaz2016absorption,goldberg2020phase}.
For $r=1$, it is known $\mathsf{AT}_{r=1}^{\textrm{Bd}}(G_N)$ is $O(N^6)$~\cite{diaz_approximating_2014} and $\mathsf{AT}_{r=1}^{\textrm{dB}}(G_N)$ is $O(N^5)$~\cite{cooper2016linear}.
For any undirected unweighted graph $G=(V,E)$ on $N$ vertices when $r>1$ and $\delta=0,1$,
we have $\mathsf{fp}_r^{\delta}(G,\{u\}) \geq \mathsf{fp}_{r=1}^{\delta}(G,\{u\}) = \Omega(N^{-2})$ for every $u\in V$, and
thus $\mathsf{FT}_{r}^{\mathrm{Bd}}(G_N)$ is $o(N^{5+\varepsilon})$ for all $\varepsilon>0$.
Similar to fixation times in neutral evolution, we prove a polynomial bound for any mixed $\delta=1/2$ process.
\begin{theorem}[Fixation time under weak selection]
  \label{thm:fixation-time-selection-half}
  If $\delta=1/2$ and $r > 1$, then the fixation time of a mutant starting in an
  arbitrary location on any graph of size $N$ is at most $N^5 \cdot \frac{r}{r-1}$.
\end{theorem}
\noindent In our notation, \Cref{thm:fixation-time-selection-half} states that for any graph $G=(V,E)$ on $N$ vertices and initial mutant location $u\in V$,
we have $\mathsf{FT}_r^{\delta=1/2}(G, \{u\}) \leq N^5 \cdot r/(r-1)$.
The proof is in~\Cref{sec:fix-time-selection-half}.

\begin{figure}
  \centering
  \includegraphics[width=\textwidth]{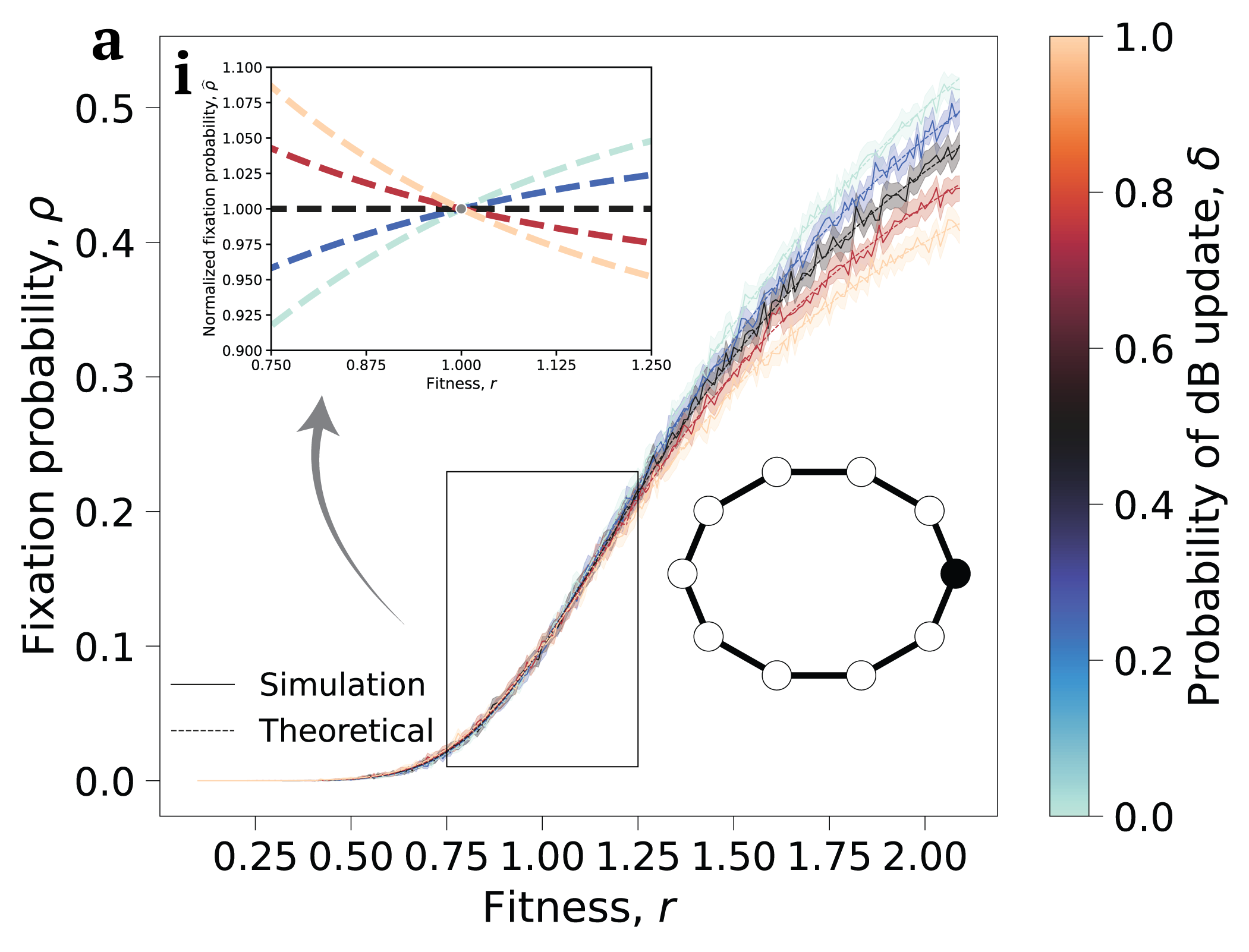}
  \caption{%
    \textbf{a,}
    Fixation probabilities on an undirected cycle with $N=10$ vertices as a function of $r$
    for $\delta=0,0.25,0.5,0.75,1$
    starting with one mutant.
    The various curves represent the various values of $\delta$.
    The solid lines represent the average of $10^4$ simulations of the mixed $\delta$-updating process, and
    the surrounding translucent regions are the $95\%$ confidence intervals.
    The dashed lines are the theoretical fixation probabilities from~\Cref{thm:cycles}.
    \textbf{i)} Zoomed in plot of the theoretical curves in the region $r\in [0.75,1.25]$,
    normalized by the fixation probability when $\delta=0.5$.
    At $r=1$, the fixation probability is $1/N$ regardless of $\delta$. This value is indicated by a dot.
    The plotted range extends below $r=1$, that is, to deleterious mutants, for which~\Cref{thm:cycles} remains valid.
    The fixation probability for the well-mixed population is equivalent to the fixation probability for a cycle under pure Bd updating (drawn in dashed light blue) for all $r$, by the Isothermal Theorem~\cite{lieberman2005evolutionary}.
  }
  \label{fig:cycle}
\end{figure}

For an undirected cycle $C_N$ on $N$ vertices, it is known for Bd that
$\mathsf{fp}_r^{\mathrm{Bd}}(C_N)=\mathsf{fp}_r^{\mathrm{Bd}}(K_N)=\frac{1-1/r}{1-1/r^N}$
for all $r\in(0,1)\cup(1,\infty)$ and $\mathsf{fp}_{r=1}^{\mathrm{Bd}}(C_N)=1/N$~\cite{lieberman2005evolutionary},
where $K_N$ denotes the complete graph on $N$ vertices with self-loops, that is, the well-mixed population;
for Bd updating the self-loops do not change the fixation probability.
For dB, it is known~\cite{kaveh2015duality,hindersin2015most,allen2020transient} that $\mathsf{fp}_{r=1}^{\mathrm{dB}}(C_N)=1/N$,
and for $r\in(0,1)\cup(1,\infty)$,
\begin{equation}
\mathsf{fp}_r^{\mathrm{dB}}(C_N) = \frac{2(r-1)}{3r-1+(r-3)r^{2-N}}.
\end{equation}
We prove a general formula for any $r>0$ and $\delta\in[0,1]$.
\begin{theorem}[Fixation probability on cycles]
  \label{thm:cycles}
  Let $r>0$ and $\delta\in[0,1]$.
  Define $F_k := rk + (N-k)$ to be the total fitness in the population with $k$ mutants.
  Define $\gamma_k$ as
  \begin{equation}
    \gamma_k =
      \begin{cases}
        \displaystyle \frac{(1-\delta)/F_k + \delta/N}{(1-\delta) r/F_k +  2r\delta/((1+r)N)}&\text{if }k=1,\\[10pt]
        \displaystyle \frac{(1-\delta)/F_k +  2\delta/((1+r)N)}{(1-\delta) r/F_k + \delta/N}&\text{if }k=N-1,\\[10pt]
        \displaystyle \frac{(1-\delta)/F_k + 2\delta/((1+r)N)}{(1-\delta) r/F_k + 2r\delta/((1+r)N)}&\text{otherwise.}\\
      \end{cases}
  \end{equation}
  Then
  \begin{equation}
     \mathsf{fp}_{r}^{\delta}(C_N) = \frac{1}{1 + \sum_{j=1}^{N-1}\prod_{k=1}^j \gamma_k}.
  \end{equation}
\end{theorem}
\noindent See Fig~\ref{fig:cycle} for a plot of the fixation probability as a function of $r$ and $\delta$.
See~\Cref{sec:cycle} for the proof.

\begin{figure}
  \centering
  \includegraphics[width=\textwidth]{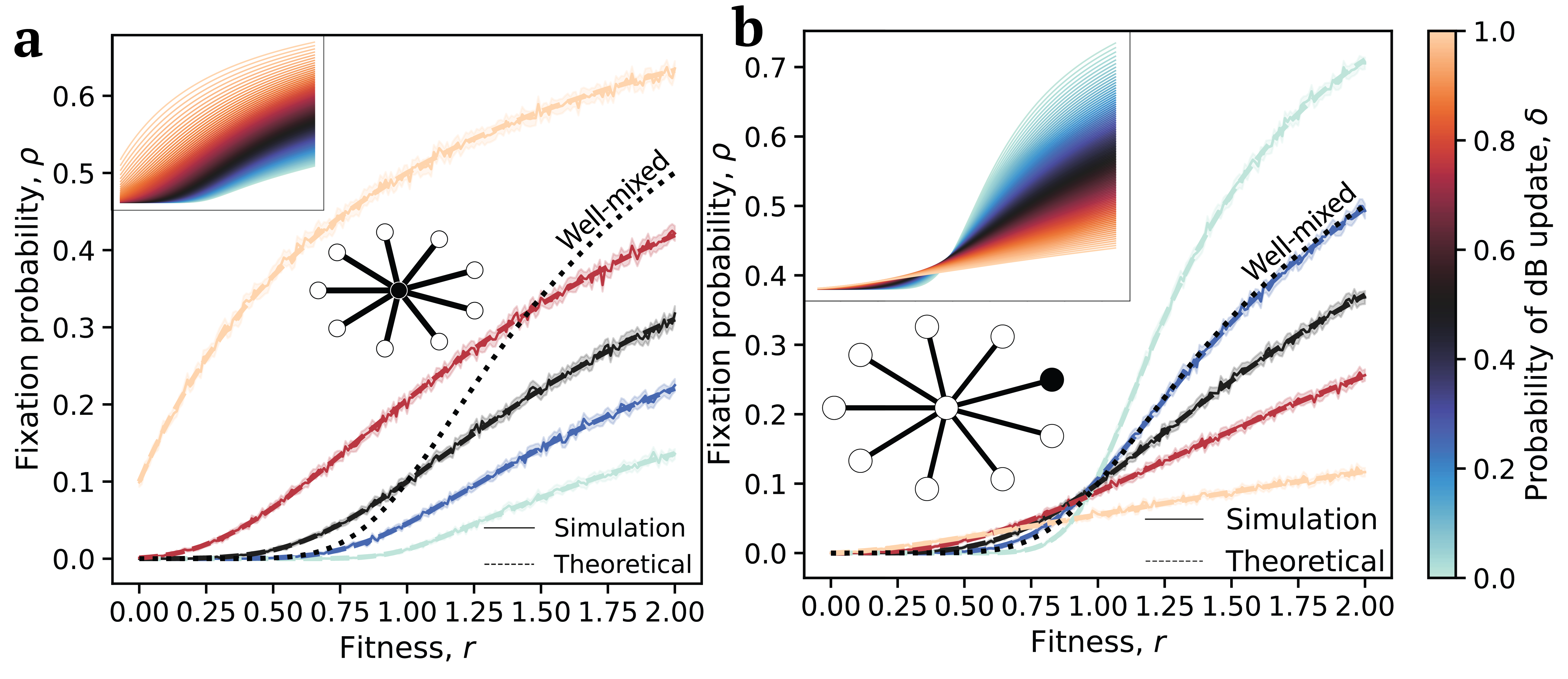}
  \caption{%
    Fixation probabilities on a star graph with $N=10$ vertices when $r\neq 1$.
    The plotted range extends below $r=1$, that is, to deleterious mutants, for which~\Cref{thm:stars-selection} remains valid.
    The various curves represent the various values of $\delta$.
    The solid lines represent the average of $10^4$ simulations of the mixed $\delta$-updating process, and
    the surrounding translucent regions are the $95\%$ confidence intervals.
    The dashed lines are the theoretical fixation probabilities from~\Cref{thm:stars-selection}.
    The dotted black line is the fixation probability for a well-mixed population~\cite{lieberman2005evolutionary}.
    \textbf{a,}
    Fixation probabilities as a function of $r$ for $\delta=0,0.25,0.5,0.75,1$
    starting with a mutant in the center.
    \textbf{b,}
    Same as a) except a mutant is initially on a leaf.
    Note that for none of the values of $\delta$ depicted do the star and well-mixed fixation probabilities
    agree for all $r$. In \textbf{b} the $\delta=0.25$ curve lies close to, but does not coincide with, the well-mixed curve.
    The inset in each panel shows the theoretical fixation probability from~\Cref{thm:stars-selection}
    on the same axes and color scale as the main panel, but for a fine sweep of $\delta$ across $[0,1]$
    rather than the five values plotted with simulations, so that the whole continuum of update rules is visible.
  }
  \label{fig:star-selection}
\end{figure}

Let $S_N$ be a star graph with one center labelled $c$ and $n:=N-1$ leaves labeled $\ell_1,\ldots,\ell_{n}$.
For Bd, it is known~\cite{broom2008analysis} that
\begin{equation}
  \mathsf{fp}_r^{\mathrm{Bd}}(S_N, \{\ell_i\})=
  \frac{nr}{nr+1}\cdot p
  \quad
  \text{and}
  \quad
  \mathsf{fp}_r^{\mathrm{Bd}}(S_N, \{c\})=
  \frac{r}{r+n}\cdot p
\end{equation}
with
\begin{equation}
  p = \frac{1}{1+\frac{n}{n+r}\sum_{j=1}^{n-1}\left(\frac{n+r}{r(nr+1)}\right)^j}.
\end{equation}
For dB, it is known~\cite{hadjichrysanthou2011evolutionary,allen2020transient} that
\begin{equation}
  \mathsf{fp}_r^{\mathrm{dB}}(S_N) = \frac{(N-1)r+1}{N(r+1)}\cdot\left(\frac1N + \frac{r}{N+2r-2}\right).
\end{equation}
We prove a general formula for any $r>0$ and $\delta\in[0,1]$.
\begin{theorem}[Fixation probability on stars]
  \label{thm:stars-selection}
  For a star $S_N$ with center $c$ and $n=N-1$ leaves, for all $\delta\in[0,1]$ and $r>0$,
  the fixation probabilities $\mathsf{fp}_r^\delta(S_N, \{c\})$ and $\mathsf{fp}_r^\delta(S_N, \{\ell_i\})$
  are given in closed form, as a product of explicit $2\times2$ transfer matrices, in~\Cref{sec:star-selection}.
  Furthermore, when $r=1$ (neutral evolution),
  these formulas recover~\Cref{thm:stars}.
\end{theorem}
\noindent The derivation is in~\Cref{sec:star-selection}.
See Fig~\ref{fig:star-selection} for a plot of the fixation probability as a function of $r$ and $\delta$ for a fixed star.

\subsubsection{Strong selection}

For pure Bd and dB, no closed form analytical formulas are known that hold for all $r$ on arbitrary graphs.
(Closed forms are, however, known for special structures such as the complete graph, the cycle, and the star, as given above.)
For Bd, there is a fully polynomial randomized approximation scheme (FPRAS)
that with confidence $1-\alpha$ outputs an estimate of $\mathsf{fp}_r^{\mathrm{Bd}}(G)$ with multiplicative error $\varepsilon$
in time polynomial in $N$, $1/\varepsilon$, and $\log(1/\alpha)$~\cite{diaz_approximating_2014,diaz2016absorption,chatterjee_et_al:LIPIcs.MFCS.2017.61}.
Furthermore, for both Bd and dB, there is a known polynomial time (with respect to $N$) algorithm that computes
$\partial\mathsf{fp}_r^{\delta}(G) / \partial r\big|_{r=1^{+}}$
for any undirected weighted graph $G$~\cite{allen2017evolutionary}.
For graphs with sufficiently large $r$ relative to the degree ratio, we prove the following bounds.
\begin{theorem}[Fixation probability lower bound under strong selection]
  \label{thm:fp-lower-bound-selection}
  For any graph $G=(V,E)$ on $N$ vertices with maximum degree $D$ and minimum degree $d$,
  for all $\delta\in[0,1]$ and $r > (D/d)^2$,
  every initial mutant set $S_0$ with $\emptyset\subsetneq S_0\subsetneq V$ satisfies
  $\mathsf{fp}_r^\delta(G, S_0) \geq |S_0|/N^2$;
  in particular $\mathsf{fp}_r^\delta(G, \{u\}) \geq 1/N^2$ for all $u\in V$.
\end{theorem}
\noindent The proof is in~\Cref{sec:strong-selection}.
\begin{theorem}[Fixation time upper bound under strong selection]
  \label{thm:fixation-time-selection}
  For any graph $G=(V,E)$ on $N\geq 4$ vertices with maximum degree $D$ and minimum degree $d$,
  for all $\delta\in[0,1]$ and $r > (D/d)^2$,
  the fixation time starting at an arbitrary location is at most $4N^7 \cdot \frac{r}{r-1}$.
\end{theorem}
\noindent See~\Cref{sec:strong-selection} for the proofs of the preceding two theorems.
Together, these two results show that when $r$ is large enough relative to the degree heterogeneity, the process behaves well.
In particular, using the bounds from \Cref{thm:fp-lower-bound-selection,thm:fixation-time-selection},
we give a simple FPRAS for $\mathsf{fp}_r^\delta(G, S_0)$.
\begin{theorem}[Fixation probability estimation algorithm]
  \label{thm:estimation-algorithm}
  For any graph $G=(V,E)$ on $N$ vertices with maximum degree $D$ and minimum degree $d$,
  for all $\delta\in[0,1]$, $r>(D/d)^2$, $\varepsilon>0$, $\alpha\in(0,1]$, and every initial mutant set $S_0\subseteq V$,
  there exists a randomized algorithm $\mathcal A$, running in time polynomial in $N$, $1/\varepsilon$, $\log(1/\alpha)$ and $r/(r-1)$,
  that with probability at least $1-\alpha$ outputs an estimate of $\mathsf{fp}_r^\delta(G, S_0)$
  within multiplicative error $\varepsilon$.
\end{theorem}
\noindent See~\Cref{sec:estimation-algo} for the proof.
Note that when $\delta=1/2$, the bound on fixation time from \Cref{thm:fixation-time-selection-half} gives an FPRAS
for \emph{all} undirected unweighted graphs with $r>1$
(see~\Cref{thm:estimation-algorithm-delta-1/2}).

\section{Discussion}
In this paper, we have studied mixed dB and Bd updating for neutral evolution and for constant selection.
At each time step we use dB updating with probability $\delta$ and Bd updating with probability $1-\delta$.
We have derived the following results, which appear in full in the preceding sections and the appendix.

The fixation probability as a function of $\delta$ can be increasing, decreasing, or possibly non-monotonic
in neutral evolution on an undirected unweighted graph.
Despite this, we find that the fixation probability is $1/N$ on any such graph when $\delta=1/2$.
Additionally, for any fixed initial mutant location, it cannot be the case that the fixation probability is greater than $1/N$ for both pure dB and pure Bd (see~\Cref{sec:not-possible}).

We give an explicit formula for the fixation probability on a bidegreed graph for any $\delta$.
Further, we prove that with a deterministic initial mutant location,
the fixation probability in neutral evolution on a bidegreed graph is the most sensitive at pure Bd or pure dB (see~\Cref{sec:sensitivity}).

Similar to the fixation probability, the fixation time as a function of $\delta$ can be increasing, decreasing, or possibly non-monotonic
in neutral evolution on an undirected unweighted graph.
We prove that the expected (conditional) fixation time on such graphs is at most some polynomial in $N$
when $\delta=1/2$ (see~\Cref{sec:fix-time-unbiased}).

For constant selection with $r>1$, we prove that when $\delta=1/2$ the fixation time is at most $N^5\cdot r/(r-1)$
on any undirected unweighted graph (see~\Cref{sec:fix-time-selection-half}).
We also prove that for graphs with $r>(D/d)^2$, the expected fixation time is at most $4N^7\cdot r/(r-1)$
for any $\delta$ (see~\Cref{sec:strong-selection}).
When $r>(D/d)^2$, we provide an efficient estimation algorithm for any $\delta$
for the fixation probability (see~\Cref{sec:estimation-algo}).

For constant selection on undirected cycles and stars, we provide tractable analytical formulas for all $\delta$
(see~\Cref{sec:cycle} and~\Cref{sec:star-selection}, respectively).

While fixation probabilities and times can have strange behaviors with respect to $\delta$,
we have shown that unbiased mixing and nearly regular graphs lead to well-behaved quantities.

There are many future directions of this work.
We highlight two.
First, one could seek to classify the graphs whose fixation probability is monotone in $\delta$, even in neutral evolution.
Second, it is tempting to ask whether the mixing parameter $\delta$ itself could be inferred for a real population.
We regard this as a speculative and challenging open direction rather than an immediate application, since
it would require empirical fixation probability (or fixation time) measurements on a population with a known structure,
and even then $\delta$ need not be uniquely identifiable from such data.
We should mention that \cite{dew2025understanding} provides an ecological context and model in which $\delta$ (not necessarily constant over time) is set by an intrinsic death rate $\Delta$ through $\delta=\Delta/(\varphi+\Delta)$, whereas the birth rate is determined independently by the type of individual.
In this setting, inference of $\delta$ is tantamount to some type of inverse problem through Bayesian analysis, though it may not be straightforward.
We leave to future work a careful investigation of whether, and under what conditions, $\delta$ can be estimated from data. \\



\enlargethispage{20pt}
\noindent\textbf{Data Availability. }{%
  Our code is available as a GitHub repository at \url{https://github.com/davidb2/mixed-updating}.
  For graph exploration, we utilized the computer program \texttt{nauty}~\cite{mckey1990nauty}.
  We also explored dynamics using \texttt{EvoLudo}~\cite{evoludo}.
} \\
\noindent \textbf{Acknowledgments. }{%
  The authors thank Josef Tkadlec for many helpful suggestions.%
} \\
\noindent\textbf{Financial Disclosure. }{%
  D.A.B. was supported by a Harvard Graduate School of Arts and Sciences Prize Fellowship.
  Y.H. and M.M. were supported by grant CNS-2107078 from the National Science Foundation (\url{https://www.nsf.gov}).
  The funders had no role in study design, data collection and analysis, decision to publish, or preparation of the manuscript.
} \\
\noindent\textbf{Competing Interests. }{%
  The authors have declared that no competing interests exist.
} \\
\noindent\textbf{Disclaimer. }{%
  A preliminary version of this work with differing content appears in an arXiv preprint for the \textit{Innovations in Theoretical Computer Science 2026} conference~\cite{brewster2025mixed}.%
}


\vskip2pc

\bibliographystyle{unsrt}
\bibliography{sample}

\appendix
\counterwithout{theorem}{section}
\counterwithout{lemma}{section}
\counterwithout{corollary}{section}
\counterwithout{definition}{section}
\setcounter{theorem}{0}
\setcounter{lemma}{0}
\setcounter{corollary}{0}
\setcounter{definition}{0}
\renewcommand{\thetheorem}{S\arabic{theorem}}
\renewcommand{\thelemma}{\arabic{lemma}}
\renewcommand{\thecorollary}{\arabic{corollary}}
\renewcommand{\thedefinition}{\arabic{definition}}

\noindent We follow the conventions of the main text. The mixed $\delta$-updating process is defined for every
$r>0$ and every $\delta\in[0,1]$, and a result stated for a \emph{neutral} mutant, or written with the subscript
$r=1$, assumes $r=1$.
Every other result states its own fitness hypothesis.

\section{Technical tools for neutral evolution}
  Our main technical tool is the theory of martingales applied to a potential function $\phi$ defined on the mutant configurations $S \subseteq V$.
  Intuitively, $\phi$ is used to reduce a high dimensional state space into a one dimensional state space to track the progress of the mutants.
  Concretely, we choose $\phi$ so that under the unbiased ($\delta=1/2$) process the expected one step change of $\phi$ is zero (making $\phi(Y_i)$ a martingale), which yields fixation probabilities via the optional stopping theorem;
  and we lower bound the expected one step \emph{squared} change of $\phi$ (its conditional variance), which yields absorption time bounds via a supermartingale argument.
  We first recall the optional stopping theorems.
  \begin{theorem}[Optional Stopping Theorem for martingales, Corollary of Theorem 17.4 of~\cite{levin2017markov}]
    \label{thm:optional-ori}
    Let $k>0$ be some constant.
    Let $(X_i)_{i \geq 0}$ be a Markov chain with state space $\mathcal X$
    such that $\{0,k\}\subseteq \mathcal X \subseteq [0,k]$.
    Let $\tau := \min \{ i : X_i \in \{0,k\} \}$.
    Suppose that $\mathbb E[\tau]<\infty$
    and
    $\mathbb{E}[X_{i+1} - X_{i} \mid X_i = x] = 0$ for all $i \geq 0$ and $x\in\mathcal X$.
    Then $\mathbb{E}[X_{\tau}] = \mathbb{E}[X_0]$.
  \end{theorem}
  \begin{theorem}[Optional Stopping Theorem for supermartingales, Corollary of Theorem 17.4 of~\cite{levin2017markov}]
    \label{thm:optional}
    Let $(Y_i)_{i\geq0}$ be a Markov chain, let $(Z_i)_{i \geq 0}$ be a sequence of random variables with $Z_i$
    determined by $Y_0,\ldots,Y_i$, and let $\tau$ be a stopping time with $\mathbb E[\tau]<\infty$.
    Suppose there is a constant $B<\infty$ with $|Z_{i+1}-Z_i|\leq B$ for all $i$,
    and $\mathbb{E}[Z_{i+1} - Z_{i} \mid Y_i] \leq 0$ on the event $\{\tau > i\}$.
    Then $\mathbb{E}[Z_{\tau}] \leq \mathbb{E}[Z_0]$.
  \end{theorem}
  \noindent Here $(Z_i)$ must have bounded increments but need not itself be bounded.
  \begin{corollary}[Fixation probability of martingale]
    \label{theorem: fixation prob}
    Let $k_2\geq k_1>0$ be some constants.
    Let $(X_i)_{i \geq 0}$ be a Markov chain with state space $\mathcal X$
    such that $\{0,k_2\}\subseteq \mathcal X \subseteq [0,k_2]$.
    Suppose $\mathbb E[X_0]=k_1$.
    Let $\tau := \min \{ i : X_i \in \{0,k_2\} \}$.
    Suppose that $\mathbb E[\tau]<\infty$
    and
    $\mathbb{E}[X_{i+1} - X_{i} \mid X_i = x] = 0$ for all $i \geq 0$ and $x\in\mathcal X$.
  Then $\mathbb{P}(X_\tau = k_2) = k_1/k_2$.
  \end{corollary}
  \begin{proof}
    Since $X_\tau\in\{0,k_2\}$, we have $\mathbb E[X_\tau]=\mathbb P(X_\tau=k_2)\cdot k_2$.
    By \Cref{thm:optional-ori}, $\mathbb E[X_\tau]=\mathbb E[X_0]=k_1$, so $\mathbb P(X_\tau=k_2)=k_1/k_2$.
  \end{proof}

  \begin{lemma}[Absorption time is finite]
    \label{lem:tau-finite}
    Let $\delta\in[0,1]$ and consider the Markov chain $(Y_i)_{i\geq 0}$ on $\Omega=2^V$ of the mixed $\delta$-updating process where $Y_i\subseteq V$ are the mutant configurations.
    Let $\phi\colon 2^V\to \mathbb{R}_{\geq0}$ with $\phi(\emptyset)=0$, and let $\tau:=\min \{ i : \phi(Y_i) \in \{0,\phi(V)\} \}$.
    Then $\mathbb E[\tau]<\infty$.
  \end{lemma}
  \begin{proof}
    Consider a contiguous block of $N$ successive states of the Markov chain
    $Y_i,\dots,Y_{i+N-1}$ with $Y_i \not\in \{\emptyset,V\}$.
    Then it must be that $|Y_i|\ge 1$.
    Consider such a contiguous block in which some $u\in Y_i$ spreads to the whole graph.
    Each step happens with probability at least $1/N^{2}$,
    so the entire sequence occurs with probability at least $p=N^{-2N}$.
    Thus, the expected number of such contiguous blocks until absorption occurs is at most $1/p<\infty$.
  \end{proof}

  \noindent Consider the Markov chain $(Y_i)_{i\geq 0}$ on $\Omega=2^V$ of the mixed $\delta$-updating process where $Y_i\subseteq V$,
  and let $\phi\colon 2^V\to\mathbb R_{\geq0}$ be a potential function on the mutant configurations.
  For an edge $(u,v)$ with $u\in S$ and $v\notin S$, we define the following quantities, each describing the expected change of the potential $\phi$ due to an update that acts along that specific edge $(u,v)$ (the $\textrm{Bd}$ and $\textrm{dB}$ versions correspond to a pure Birth-death and a pure death-Birth step, respectively, and the $\delta$ version to the mixed step):
  \begin{align}
    \psi_{u, v}^{\textrm{Bd}}(S) &:= \frac 1{N} \left(\frac 1 {\deg(u)} (\phi(S\cup \{v\})-\phi(S)) - \frac 1 {\deg(v)} (\phi(S)-\phi(S\setminus \{u\})) \right)\\
    \psi_{u, v}^{\textrm{dB}}(S) &:= \frac 1{N} \left(\frac 1 {\deg(v)}(\phi(S\cup \{v\})-\phi(S)) - \frac 1 {\deg(u)} (\phi(S)-\phi(S\setminus \{u\}))  \right) \\
    \psi_{u, v}^{\delta} (S) &:= (1-\delta) \psi_{u, v}^{\textrm{Bd}}(S) + \delta \psi_{u, v}^{\textrm{dB}}(S).
  \end{align}
  Summing over all crossing edges recovers the total expected one step change, $\mathbb E[\phi(Y_{i+1})-\phi(Y_i)\mid Y_i=S]=\sum_{u\in S,\,v\notin S,\,(u,v)\in E}\psi_{u, v}^{\delta} (S)$.
  In particular, for a single crossing edge $\psi_{u, v}^{\delta} (S)$ is exactly the contribution of that edge to $\mathbb E[\phi(Y_{i+1})-\phi(Y_i)\mid Y_i=S]$.

\section{Additivity of fixation probability}
\label{sec:additivity}
\begin{lemma}[Additivity in neutral evolution]
  \label{claim:fp-additive}
  Let $G=(V,E)$ be a connected undirected graph.
  For any $T\subseteq S \subseteq V$ we have
  \begin{equation}
    \mathsf{fp}_{r=1}^\delta(G, S) = \mathsf{fp}_{r=1}^\delta(G, T)+\mathsf{fp}_{r=1}^\delta(G, S\setminus T).
  \end{equation}
\end{lemma}
\begin{proof}
  Consider the variant of the mixed $\delta$-updating process where every vertex starts with a unique type.
  The process will evolve as usual until a single type takes over the whole graph.
  A mutant currently occupying subset $S$ fixates in the mixed $\delta$-updating process if and only if one of the vertices in the subset takes over in the altered process,
  which in turn equals the fixation probability with only that vertex by viewing all other types as wild types. 
\end{proof}

\section{Proof of~\Cref{thm:neutral-evolution}}
\label{sec:proof-neutral-evolution}
\begin{proof}
  We consider the Markov chain $(Y_i)_{i\geq 0}$ on $\Omega=2^V$ of the mixed $1/2$-updating process where $Y_i\subseteq V$.
  Let $S\subseteq V$.
  We use the potential function $\phi(S)=|S|$.
  Let $\tau:=\min \{ i : \phi(Y_i) \in \{0,N\} \}$.
  By~\Cref{lem:tau-finite}, $\tau$ has finite expectation.
  For this potential, each of the two increments appearing in the definitions of $\psi_{u,v}^{\text{Bd}}$ and $\psi_{u,v}^{\text{dB}}$ equals $1$. Indeed,
  for every crossing edge $(u,v)$ with $u\in S$ and $v\notin S$ we have $\phi(S\cup\{v\})-\phi(S)=1$ and $\phi(S)-\phi(S\setminus\{u\})=1$.
  Substituting into the definitions gives
  \begin{equation}
    \psi_{u,v}^{\text{Bd}}(S)=\frac1N\left(\frac{1}{\deg(u)}-\frac{1}{\deg(v)}\right)
    =-\frac1N\left(\frac{1}{\deg(v)}-\frac{1}{\deg(u)}\right)=-\psi_{u,v}^{\text{dB}}(S),
  \end{equation}
  so that $\psi_{u,v}^{\delta=1/2}(S)=\tfrac12\bigl(\psi_{u,v}^{\text{Bd}}(S)+\psi_{u,v}^{\text{dB}}(S)\bigr)=0$ for every crossing edge.
  Summing over all crossing edges yields
  $\mathbb{E}[\phi(Y_{i+1})-\phi(Y_i)\mid Y_i=S]=\sum_{u\in S,\,v\notin S,\,(u,v)\in E}\psi_{u,v}^{\delta=1/2}(S)=0$.
  Then we can apply~\Cref{theorem: fixation prob} to yield the desired result.
\end{proof}

\section{Proof that it is not possible for both pure processes to simultaneously have fixation probabilities greater than $1/N$}
\label{sec:not-possible}
\begin{lemma}
  \label{thm:both-pure-larger-than-unbiased-not-possible}
  Let $G=(V,E)$ and $u\in V$.
  Then it must be the case that
  \begin{equation}
    \min\{\mathsf{fp}_{r=1}^{\mathrm{dB}}(G, \{u\}), \mathsf{fp}_{r=1}^{\mathrm{Bd}}(G, \{u\})\} \leq 1/N.
  \end{equation}
\end{lemma}
\begin{proof}
  From~\cite{broom2010two,svoboda2024amplifiers},
  we know that 
  \begin{equation}
    \mathsf{fp}_{r=1}^{\mathrm{Bd}}(G, \{u\}) \propto 1/\deg(u)
    \quad\text{and}\quad
    \mathsf{fp}_{r=1}^{\mathrm{dB}}(G, \{u\}) \propto \deg(u).
  \end{equation}

  Suppose that $\mathsf{fp}_{r=1}^{\mathrm{dB}}(G, \{u\})>1/N$.
  Then it must be that $\deg(u)>D_{\textrm{AM}}$ where we take
  ${D_{\textrm{AM}}:=\frac1N\cdot\sum_{v\in V}\deg(v)}$
  to be the arithmetic mean of the degrees of the vertices in $G$.
  Since the arithmetic mean is at least the harmonic mean of a sequence,
  we have that $\deg(u)\geq D_{\textrm{HM}}$ where ${D_\textrm{HM}:=\frac{N}{\sum_{v\in V}1/\deg(v)}}$ is the 
  harmonic mean of the degrees of the vertices in $G$.
  Thus
  \begin{equation}
    1/\deg(u) \leq 1/D_\textrm{HM}
    = \frac1N\sum_{v\in V} \frac1{\deg(v)}.
  \end{equation}
  This means that $\mathsf{fp}_{r=1}^{\mathrm{Bd}}(G, \{u\})\leq 1/N$.

  On the other hand, suppose that $\mathsf{fp}_{r=1}^{\mathrm{Bd}}(G, \{u\})>1/N$.
  Then it must be that $\deg(u)<D_{\textrm{HM}}$.
  Since $D_{\textrm{HM}}\leq D_{\textrm{AM}}$,
  we have that $\deg(u)\leq D_{\textrm{AM}}$.
  Thus $\mathsf{fp}_{r=1}^{\mathrm{dB}}(G, \{u\})\leq 1/N$.
\end{proof}

\section{Proof of~\Cref{thm:bidegreed}}
\label{sec:bidegreed-main}
\begin{proof}[Proof of~\Cref{thm:bidegreed}]
  We denote the function $B(d_i)$ as $1$ if $i=1$ and $\frac{(1-\delta)d_1+\delta d_2}{(1-\delta)d_2+\delta d_1}$ if $i=2$.
  We consider the Markov chain $(Y_i)_{i\geq 0}$ on $\Omega=2^V$ of the mixed $\delta$-updating process where $Y_i\subseteq V$. 
  We use the potential function $\phi\colon 2^V\to\mathbb R_{\geq0}$ such that
  $\phi(S)=\sum_{v\in S}B(\deg(v))$.
  If $u\in S$ and $v\not\in S$,
  then
  \begin{equation}
    \psi_{u,v}^{\delta}(S)
    = \frac{\delta}{N}\left(\frac{B(\deg(v))}{\deg(v)}-\frac{B(\deg(u))}{\deg(u)}\right)
    + \frac{1-\delta}{N}\left(\frac{B(\deg(v))}{\deg(u)}-\frac{B(\deg(u))}{\deg(v)}\right).
  \end{equation}
  \begin{itemize}
    \item If $\deg(u)=\deg(v)$, then clearly $\psi_{u,v}^{\delta}(S)=0$.
    \item Otherwise, if $\deg(u)<\deg(v)$ then
    \begin{align}
      \psi_{u,v}^{\delta}(S)
      &=\frac{1}{N} \cdot \left[ \delta\left(\frac{B(d_2)}{d_2}-\frac{1}{d_1}\right)
      +                          (1-\delta)\left(\frac{B(d_2)}{d_1}-\frac{1}{d_2} \right) \right] \\
      &= \frac{1}{Nd_1d_2}\cdot\left[ \delta(B(d_2)d_1-d_2) + (1-\delta)(B(d_2)d_2-d_1) \right] \\
      &= \frac{1}{Nd_1d_2}\cdot\left[ B(d_2)((1-\delta)d_2+\delta d_1) - ((1-\delta)d_1 + \delta d_2) \right] \\
      &=0.
    \end{align}
    \item Due to symmetry, the case of $\deg(u)>\deg(v)$ follows from the same argument as the preceding case.
  \end{itemize}
  Thus in any case, $\mathbb E[\phi(Y_{i+1})-\phi(Y_i)\mid Y_i=S]=0$.
  Applying~\Cref{theorem: fixation prob} with $k_1=B(\deg(u))$ and $k_2=\sum_{v\in V}B(\deg(v))$
  completes the proof.
\end{proof}

\subsection{Sensitivity of fixation probabilities on bidegreed graphs}
\label{sec:sensitivity}
\begin{theorem}
\label{thm:sensitivity}
Let $G=(V,E)$ be a $(d_1,d_2)$-bidegreed graph on $N\geq 1$ vertices with $d_1\le d_2$.
Let $n_1$ be the number of vertices of degree $d_1$
and $n_2$ the number of vertices of degree $d_2$
such that $N=n_1+n_2$ is the total number of vertices.
When $G$ is regular, so that $d_1=d_2$, we adopt the counting convention $n_1=N$ and $n_2=0$,
so that $N=n_1+n_2$ continues to hold.
For $\delta\in[0,1]$, let 
\begin{equation}
\rho_1(\delta) := \mathsf{fp}_{r=1}^\delta(G,\{u\})\quad\text{for any }u\in V\text{ with }\deg(u)=d_1,
\end{equation}
\begin{equation}
\rho_2(\delta) := \mathsf{fp}_{r=1}^\delta(G,\{u\})\quad\text{for any }u\in V\text{ with }\deg(u)=d_2.
\end{equation}
Then for all $\delta\in[0,1]$, the values $\rho'_1(\delta)$ and $\rho'_2(\delta)$ exist and are given by
$
\rho_1'(\delta)
=-n_2 \cdot Q(\delta)
$
and
$
\rho_2'(\delta)
=n_1 \cdot Q(\delta)
$
where
\begin{equation}
Q(\delta)
:=\frac{(d_2-d_1)(d_1+d_2)}{D(\delta)^2}
\end{equation}
and
\begin{equation}
D(\delta):=d_1 n_2 + d_2 n_1 + \delta(d_1-d_2)(n_1-n_2).
\end{equation}
If $d_1<d_2$, then for all $\delta\in[0,1]$ one has $\rho_1'(\delta)<0$ and $\rho_2'(\delta)>0$. 
Otherwise, $\rho_1'(\delta)=\rho_2'(\delta)=0$ for all $\delta$.
\end{theorem}

\begin{proof}
We denote the function $B(d_i)$ as $1$ if $i=1$ and $\frac{(1-\delta)d_1+\delta d_2}{(1-\delta)d_2+\delta d_1}$ if $i=2$.
By~\Cref{thm:bidegreed}, for any vertex $u\in V$ with degree $d_1$ or $d_2$ we have
\begin{equation}
\mathsf{fp}_{r=1}^\delta(G,\{u\})=
\frac{B(\deg(u))}{\sum_{v\in V} B(\deg(v))}.
\end{equation}
Let
\begin{equation}
S(\delta) := \sum_{v\in V} B(\deg(v)) = n_1 + n_2 B_2(\delta)
\quad\text{where}
\quad B_2(\delta) :=  \frac{(1-\delta)d_1+\delta d_2}{(1-\delta)d_2+\delta d_1}.
\end{equation}
Then
$\rho_1(\delta) = \frac{1}{S(\delta)}$ and
$\rho_2(\delta) = \frac{B_2(\delta)}{S(\delta)}$.

We first compute $B_2'(\delta)$.
Write $B_2(\delta)=A(\delta)/C(\delta)$ where
\begin{equation}
A(\delta) = (1-\delta)d_1+\delta d_2 = d_1 + \delta(d_2-d_1),
\end{equation}
\begin{equation}
C(\delta) = (1-\delta)d_2+\delta d_1 = d_2 + \delta(d_1-d_2).
\end{equation}
Then
$A'(\delta) = d_2-d_1$ and $C'(\delta) = -(d_2-d_1)$.
By the quotient rule, and since $A(\delta)+C(\delta)=d_1+d_2$,
\begin{equation}
B_2'(\delta)
=
\frac{A' C - A C'}{C^2}
=
\frac{(d_2-d_1)\bigl(C(\delta)+A(\delta)\bigr)}{C^2}
=
\frac{(d_2-d_1)(d_1+d_2)}{\bigl((1-\delta)d_2+\delta d_1\bigr)^2}.
\end{equation}
In particular $B_2'(\delta)$ has the same sign as $(d_2-d_1)$.

Next we differentiate $S(\delta)$ to get
$S'(\delta) = n_2 B_2'(\delta)$.
Hence
\begin{equation}
\rho_1'(\delta) = -\frac{S'(\delta)}{S(\delta)^2} = -\frac{n_2 B_2'(\delta)}{S(\delta)^2}.
\end{equation}
Substituting the explicit form of $B_2'(\delta)$ and clearing denominators, one finds
\begin{equation}
\rho_1'(\delta)
=-\frac{n_2(d_2-d_1)(d_1+d_2)}{D(\delta)^2}.
\end{equation}
A direct computation gives $D(\delta)=S(\delta)C(\delta)$ exactly,
so $D(\delta)>0$ for all $\delta\in[0,1]$
and the sign of $\rho_1'(\delta)$ agrees with the sign of $(d_1-d_2)$ which is non-positive.

For $\rho_2$, differentiating $\rho_2(\delta) = B_2(\delta)/S(\delta)$
and using the quotient rule gives
\begin{equation}
\rho_2'(\delta)
=\frac{B_2'S - B_2S'}{S^2}
=\frac{B_2'(\delta)\bigl(S(\delta) - n_2B_2(\delta)\bigr)}{S(\delta)^2}
=\frac{B_2'(\delta) n_1}{S(\delta)^2}.
\end{equation}
Substituting $B_2'(\delta)$ again and simplifying yields
\begin{equation}
\rho_2'(\delta)
=
\frac{n_1(d_2-d_1)(d_1+d_2)}{D(\delta)^2}.
\end{equation}

Thus when $d_1<d_2$ we have $\rho_1'(\delta)<0$ and $\rho_2'(\delta)>0$ for all $\delta$.
Therefore $\rho_1$ is strictly decreasing and $\rho_2$ strictly increasing on $[0,1]$ when $d_1<d_2$,
and the endpoint maximum and minimum follow immediately.

When $d_1=d_2$, the factor $(d_2-d_1)$ vanishes, so $\rho_1'(\delta)=\rho_2'(\delta)=0$ for all $\delta\in[0,1]$.
Therefore $\rho_1$ and $\rho_2$ are constant on $[0,1]$.

\end{proof}

The three claims stated after~\Cref{thm:bidegreed-behavior} in the main text follow.
\begin{corollary}[Endpoints of the sensitivity]
  \label{cor:sensitivity-endpoints}
  Let $G$ be a $(d_1,d_2)$-bidegreed graph and let $s_i(\delta):=|\rho_i'(\delta)|$ for $i=1,2$.
  If $d_1<d_2$ and $n_1>n_2$, then $s_i$ attains its maximum at $\delta=1$ and its minimum at $\delta=0$.
  If $d_1<d_2$ and $n_1<n_2$, then $s_i$ attains its maximum at $\delta=0$ and its minimum at $\delta=1$.
  If $n_1=n_2$ or $d_1=d_2$, then $s_i$ is constant in $\delta$.
\end{corollary}
\begin{proof}
For the magnitudes $|\rho_1'(\delta)|$ and $|\rho_2'(\delta)|$ note that
by~\Cref{thm:sensitivity},
\begin{equation}
|\rho_1'(\delta)| = \frac{n_2(d_2-d_1)(d_1+d_2)}{D(\delta)^2},\qquad
|\rho_2'(\delta)| = \frac{n_1(d_2-d_1)(d_1+d_2)}{D(\delta)^2},
\end{equation}
so the $\delta$-dependence is entirely in the factor $1/D(\delta)^2$.
Notice that $D$ is strictly monotone in $\delta$ unless $n_1=n_2$ or $d_1=d_2$,
in which case $D$ is constant.
If $n_1>n_2$ then $n_1-n_2>0$ and $d_1-d_2<0$,
so $D$ is strictly decreasing on $[0,1]$,
hence $1/D(\delta)^2$ and therefore $|\rho_i'(\delta)|$ is strictly increasing in $\delta$,
so its maximum occurs at $\delta=1$ and minimum at $\delta=0$.
If $n_1<n_2$ then $n_1-n_2<0$ and $D$ is strictly increasing,
so $|\rho_i'(\delta)|$ is strictly decreasing,
with maximum at $\delta=0$ and minimum at $\delta=1$.
If $n_1=n_2$ then $D(\delta)$ is independent of $\delta$ and hence $|\rho_i'(\delta)|$ is constant in $\delta$, and
the same holds when $d_1=d_2$, since then $\rho_1'\equiv\rho_2'\equiv0$ by~\Cref{thm:sensitivity}.
\end{proof}

\section{Fixation time}

\subsection{Proof of~\Cref{thm:fixation-time-unbiased-neutral}}
\label{sec:fix-time-unbiased}
\begin{proof}
  We consider the Markov chain $(Y_i)_{i\geq 0}$ on $\Omega=2^V$ of the mixed $1/2$-updating process where $Y_i\subseteq V$.
  Let $\phi(S) = |S|$ and $\tau:=\min \{ i : \phi(Y_i) \in \{0,N\} \}$.
  We want to use \Cref{thm:optional}.
  By~\Cref{lem:tau-finite}, $\tau$ has finite expectation.

  We now bound $\mathbb E[(\phi(Y_{i+1})-\phi(Y_i))^2 \mid Y_i=S]$ for all $S\notin\{\emptyset,V\}$.
  For any $S \notin \{\emptyset, V\}$ and edge $(u,v) \in E$
  such that either (i) $u\in S$ and $v\not\in S$ or (ii) $u\not\in S$ and $v\in S$,
  with probability $1/N^2$ the step happens along $(u,v)$ and $\phi$ changes by $\pm 1$.
  So $\mathbb E[(\phi(Y_{i+1})-\phi(Y_i))^2 \mid Y_i=S] \geq 1/N^2$.

  Define $g(x)=x(N-x)$ and
  $Z_i=g(\phi(Y_i))+\frac{i}{N^2}$.
  For any $x$ and $\Delta\in\{-1,0,1\}$,
  \begin{equation}
  g(x+\Delta)-g(x)=(N-2x)\Delta-\Delta^2.
  \end{equation}
  Exactly as in the proof of~\Cref{thm:neutral-evolution} (with $\phi(S)=|S|$, so that both bracketed increments equal $1$),
  every crossing edge satisfies $\psi_{u,v}^{\delta=1/2}(S)=0$, so
  summing over crossing edges gives $\mathbb{E}[\phi(Y_{i+1})-\phi(Y_i)\mid Y_i=S]=0$, and we get
  \begin{align}
  \mathbb{E}[Z_{i+1}-Z_i\mid Y_i = S]
  &=\mathbb{E}[g(\phi(Y_{i+1}))-g(\phi(Y_i))\mid Y_i=S]+\frac{1}{N^2} \\
  &=-\mathbb{E}[(\phi(Y_{i+1})-\phi(Y_i))^2\mid Y_i=S]+\frac{1}{N^2}\leq 0,
  \end{align}
  so $(Z_{i})$ has non-positive drift up to $\tau$, with one step differences bounded by $N+1/N^2$.
  Thus, applying \Cref{thm:optional} at the absorption time $\tau$,
  we get
  $\mathbb{E}[Z_\tau]\leq \mathbb{E}[Z_0]=g(\phi(Y_0))$.
  Since $\phi(Y_\tau)\in\{0,N\}$, we have
  $g\left(\phi(Y_\tau)\right)=0$ and
  $g(\phi(Y_0))=\phi(Y_0)(N-\phi(Y_0)) \le N^2/4$.
  Thus when $Y_0=S_0$,
  \begin{equation}
    \mathsf{AT}_{r=1}^{\delta=1/2}(G, S_0) = \mathbb E[\tau] \le N^2 \cdot g\left(\phi(S_0)\right) \le \frac{N^4}{4}.
  \end{equation}
  To convert this absorption time bound into a fixation time bound, recall the law of total expectation inequality
  $\mathsf{FT}_{r=1}^{\delta}(G,S_0)\leq \mathsf{AT}_{r=1}^{\delta}(G,S_0)/\mathsf{fp}_{r=1}^{\delta}(G,S_0)$.
  By~\Cref{thm:neutral-evolution}, $\mathsf{fp}_{r=1}^{\delta=1/2}(G,\{u\})=1/N$ for every starting vertex $u$, so
  \begin{equation}
    \mathsf{FT}_{r=1}^{\delta=1/2}(G, \{u\})
    \leq \frac{\mathsf{AT}_{r=1}^{\delta=1/2}(G, \{u\})}{\mathsf{fp}_{r=1}^{\delta=1/2}(G, \{u\})}
    \leq \frac{N^4/4}{1/N} = \frac{N^5}{4}. \qedhere
  \end{equation}
\end{proof}

\subsection{Proof of~\Cref{thm:fixation-time-bidegreed}}
\label{sec:fix-time-bidegreed}
\begin{proof}
  The proof is similar to that of~\Cref{thm:fixation-time-unbiased-neutral} but more general.
  The only difference is that we use a different potential function and can bound the expected squared potential
  change more carefully using the bidegreed property.

  We denote the function $B(d_i)$ as $1$ if $i=1$ and $\frac{(1-\delta)d_1+\delta d_2}{(1-\delta)d_2+\delta d_1}$ if $i=2$.
  Let $S(\delta) := \sum_{u\in V} B(\deg(u)) = n_1 + n_2 B(d_2)$.
  We consider the Markov chain $(Y_i)_{i\geq 0}$ on $\Omega=2^V$ of the mixed $\delta$-updating process where $Y_i\subseteq V$.
  Let $\phi(S) = \sum_{u\in S}B(\deg(u))$ and $\tau:=\min \{ i : \phi(Y_i) \in \{0,\phi(V)\} \}$.
  By~\Cref{lem:tau-finite}, $\tau$ has finite expectation.

  We now bound $\mathbb E[(\phi(Y_{i+1})-\phi(Y_i))^2 \mid Y_i=S]$ for all $S\notin\{\emptyset,V\}$.
  For any $S \notin \{\emptyset, V\}$ and edge $(u,v) \in E$
  such that either (i) $u\in S$ and $v\not\in S$ or (ii) $u\not\in S$ and $v\in S$,
  with probability $1/N^2$ the step happens along $(u,v)$ and $\phi$ changes by at least $d_1/d_2$ in absolute value. Indeed,
  the change is $\pm B(\deg(w))$ for the single vertex $w$ that switches type, and $B(d_1)=1\geq d_1/d_2$ while $B(d_2)\geq d_1/d_2$ for all $\delta$.
  So $\mathbb E[(\phi(Y_{i+1})-\phi(Y_i))^2 \mid Y_i=S] \geq \left(\frac{d_1/d_2}{N}\right)^2$.

  Define $g(x)=x(\phi(V)-x)$ and
  $Z_i=g(\phi(Y_i))+i \cdot \left(\frac{d_1/d_2}{N}\right)^2$.
  For $\Delta_i:=\phi(Y_{i+1})-\phi(Y_i)$,
  \begin{equation}
  g(\phi(Y_{i+1}))-g(\phi(Y_i))=(\phi(V)-2\phi(Y_i))\cdot \Delta_i-\Delta_i^2.
  \end{equation}
  Therefore using the fact from the proof of~\Cref{thm:bidegreed}
  that $\mathbb{E}[\phi(Y_{i+1})-\phi(Y_i)\mid Y_i=S]=0$, we get
  \begin{align}
  \mathbb{E}[Z_{i+1}-Z_i\mid Y_i = S]
  &=\mathbb{E}[g(\phi(Y_{i+1}))-g(\phi(Y_i))\mid Y_i=S]+\left(\frac{d_1/d_2}{N}\right)^2\ \\
  &=-\mathbb{E}[(\phi(Y_{i+1})-\phi(Y_i))^2\mid Y_i=S]+\left(\frac{d_1/d_2}{N}\right)^2\leq 0,
  \end{align}
  so $(Z_{i})$ has non-positive drift up to $\tau$, with uniformly bounded one step differences.
  Thus, applying \Cref{thm:optional} at the absorption time $\tau$,
  we get
  $\mathbb{E}[Z_\tau]\leq \mathbb{E}[Z_0]=g(\phi(Y_0))$.
  Since $\phi(Y_\tau)\in\{0,\phi(V)\}$, we have
  $g\left(\phi(Y_\tau)\right)=0$ and
  $g(\phi(Y_0))=\phi(Y_0)(\phi(V)-\phi(Y_0)) \leq \phi(V)^2/4$.
  Note that $\phi(V)=S(\delta)$.
  Thus when $Y_0=S_0$,
  \begin{equation}
    \mathsf{AT}_{r=1}^{\delta}(G, S_0)
    = \mathbb E[\tau]
    \le N^2 \cdot (d_2/d_1)^2 \cdot g\left(\phi(S_0)\right)
    \le \frac{S(\delta)^2\cdot (d_2/d_1)^2\cdot N^2}{4}.
  \end{equation}
  We now convert this into a fixation time bound using
  $\mathsf{FT}_{r=1}^{\delta}(G,S_0)\leq \mathsf{AT}_{r=1}^{\delta}(G,S_0)/\mathsf{fp}_{r=1}^{\delta}(G,S_0)$.
  Note that $S(\delta)=\phi(V)=n_1+n_2 B(d_2)\leq N\cdot (d_2/d_1)$, since $B(d_2)\leq d_2/d_1$ for all $\delta$.
  By~\Cref{thm:bidegreed}, a neutral mutant on a degree $d_1$ vertex has fixation probability $1/S(\delta)$, so
  \begin{equation}
    \mathsf{FT}_{r=1}^{\delta}(G, \{u\})
    \leq \mathsf{AT}_{r=1}^{\delta}(G, \{u\})\cdot S(\delta)
    \leq \frac{S(\delta)^3\, (d_2/d_1)^2\, N^2}{4}
    \leq \frac{N^5\,(d_2/d_1)^5}{4}.
  \end{equation}
  A neutral mutant on a degree $d_2$ vertex has fixation probability $B(d_2)/S(\delta)\geq (d_1/d_2)/S(\delta)$
  (since $B(d_2)\geq d_1/d_2$ for all $\delta$), so
  \begin{equation}
    \mathsf{FT}_{r=1}^{\delta}(G, \{u\})
    \leq \mathsf{AT}_{r=1}^{\delta}(G, \{u\})\cdot \frac{S(\delta)\, d_2}{d_1}
    \leq \frac{N^5\,(d_2/d_1)^6}{4}. \qedhere
  \end{equation}
\end{proof}

\section{Technical Tools for Constant Selection}

The following theorem is proven in~\cite{diaz_approximating_2014}.
\begin{theorem}[Absorption time, negative drift]
  \label{thm:negative-drift}
  Let $(Y_i)_{i \geq 0}$ be a Markov chain with state space $\Omega$, where $Y_0$ is chosen from some set $I \subseteq \Omega$.
  If there are constants $k_1, k_2 > 0$ and a non-negative function $\phi : \Omega \to \mathbb{R}$ such that
  \begin{itemize}
    \item $\phi(S) = 0$ for some $S \in \Omega$,
    \item $\phi(S) \leq k_1$ for all $S \in I$, and
    \item $\mathbb{E}[\phi(Y_i) - \phi(Y_{i+1}) \mid Y_i = S] \geq k_2$ for all $i \geq 0$ and all $S$ with $\phi(S) > 0$,
  \end{itemize}
  then $\mathbb{E}[\tau] \leq k_1 / k_2$, where $\tau = \min \{ i : \phi(Y_i) = 0 \}$.
\end{theorem}

\begin{corollary}[Absorption time, positive drift]
  \label{theorem: absorption time advantage}
  Let $(Y_i)_{i \geq 0}$ be a Markov chain with state space $\Omega$, where $Y_0$ is chosen from some set $I \subseteq \Omega$.
  If there are constants $k_1, k_2 > 0$ and a non-negative function $\phi : \Omega \to \mathbb{R}$ such that
  \begin{itemize}
    \item $\phi(T_0) = 0$ for some $T_0 \in \Omega$, and $\phi(T_1) = k_1$ for some $T_1 \in \Omega$,
    \item $\phi(S) \leq k_1$ for all $S \in I$, and
    \item $\mathbb{E}[\phi(Y_{i+1}) - \phi(Y_{i}) \mid Y_i = S] \geq k_2$ for all $i \geq 0$ and all $S$ with $\phi(S) \notin \{0, k_1\}$,
  \end{itemize}
  then $\mathbb{E}[\tau] \leq k_1 / k_2$, where $\tau = \min \{ i: \phi(Y_i) = 0 \vee \phi(Y_i) = k_1 \}$.
\end{corollary}
\begin{proof}
  Consider the process $Y_i'$ which behaves identically to $Y_i$
  except that if $\phi(Y'_i)=0$, then $\phi(Y'_{i+1})=S$ for some $S$ with $\phi(S)=k_1$.
  Let $\phi'(S)=k_1-\phi(S)$.
  Then $\min\{i: \phi'(Y_i')=0\} = \min\{i: \phi(Y_i')=k_1\} \geq \tau$.
  Now $\phi'(Y_i')$ satisfies the conditions in \Cref{thm:negative-drift} with the same parameters.
  Applying \Cref{thm:negative-drift} gives the claim.
\end{proof}

\begin{theorem}[Fixation probability, lower bound]
  \label{theorem: fixation prob selection}
  Let $(Y_i)_{i \geq 0}$ be a Markov chain with state space $\Omega$, where $Y_0$ is chosen from some set $I \subseteq \Omega$.
  Suppose there are constants $k_2 \geq k_1 > 0$ and a non-negative function $\phi\colon \Omega \to \mathbb{R}$ such that
  \begin{itemize}
    \item $\phi(T_0) = 0$ for some $T_0 \in \Omega$, and $\phi(T_1) = k_2$ for some $T_1 \in \Omega$,
    \item $\phi(Y_0) \ge k_1$,
    \item $\phi(S) \le k_2$ for all $S \in I$,
    \item $\mathbb{E}[\phi(Y_{i+1}) - \phi(Y_{i}) \mid Y_i = S] \ge 0$ for all $i \ge 0$ and all $S$ with $\phi(S) \notin \{0, k_2\}$,
    \item $\mathbb{E}[\tau] < \infty$, where $\tau := \min \{ i \ge 0 : \phi(Y_i) \in \{0, k_2\} \}$.
  \end{itemize}
  Then $\mathbb{P}(\phi(Y_\tau) = k_2) \geq k_1/k_2$.
\end{theorem}
\begin{proof}
  By the claimed properties, the stopped sequence $(\phi(Y_{i\wedge\tau}))_{i\geq0}$ is a bounded submartingale.
  By the optional stopping theorem,
  $\mathbb P(\phi(Y_\tau)=k_2) \cdot k_2 = \mathbb E[\phi(Y_\tau)]\geq \mathbb E[\phi(Y_0)] \geq k_1$.
\end{proof}

A particular notion will be useful in our proofs, describing the expected change due to a specific edge $(u,v)$:
\begin{align}
  \psi_{u, v}^{\textrm{Bd}}(S) &= \frac{1}{w(S)} \left(\frac{r}{\deg(u)}(\phi(S\cup \{v\})-\phi(S)) - \frac{1}{\deg(v)}(\phi(S)-\phi(S\setminus \{u\})) \right),\\
  \psi_{u, v}^{\textrm{dB}}(S) &= \frac{1}{N} \left(\frac{r}{w_v(S)}(\phi(S\cup \{v\})-\phi(S)) - \frac{1}{w_u(S)}(\phi(S)-\phi(S\setminus \{u\})) \right),\\
  \psi_{u, v}^{\delta}(S) &= (1-\delta) \psi_{u, v}^{\textrm{Bd}}(S) + \delta \psi_{u, v}^{\textrm{dB}}(S),
\end{align}
where $w(S) = N + (r-1)|S|$ is the total fitness and $w_u(S) = \sum_{v:(u,v)\in E} f_S(v)$ is the fitness sum of neighbors of $u$.

\section{Proof of \Cref{thm:fixation-time-selection-half}}
\label{sec:fix-time-selection-half}

\begin{theorem}
  \label{thm:abs-time-advantageous-half}
  Let $G=(V,E)$ be an undirected unweighted graph on $N$ vertices and let $r>1$.
  With $\phi(S) = |S|$ we have
  \begin{equation}
    \psi^{\delta=1/2}_{u, v}(S) \ge \frac{r-1}{rN^3}
  \end{equation}
  for every $S$ with $\emptyset\subsetneq S\subsetneq V$ and every crossing edge $(u,v)\in E$ with $u\in S$, $v\notin S$, and
  \begin{equation}
    \mathsf{AT}^{\delta=1/2}_{r}(G)\leq \frac{r}{r-1}\cdot N^4.
  \end{equation}
\end{theorem}
\begin{proof}
  The potential function is $\phi(S) = |S|$. Thus
  \begin{align}
    \psi_{u, v}^{\delta=1/2}(S) &= \frac{1}{2} \left[\frac{1}{w(S)} \left(\frac{r}{\deg(u)} - \frac{1}{\deg(v)} \right)+ \frac{1}{N} \left(\frac{r}{w_v(S)} - \frac{1}{w_u(S)} \right)\right].
  \end{align}
  For all $S \neq V$ and $S \neq \emptyset$, we have $N-1+r \leq w(S) \leq (N-1)r+1$, therefore
  \begin{align}
    \frac{1}{w(S)}\frac{r}{\deg(u)} - \frac{1}{N}\frac{1}{w_u(S)}
    &\geq \frac{r}{(r(N-1)+1)\deg(u)} - \frac{1}{N\deg(u)}
    = \frac{1}{\deg(u)}\frac{1}{N}\frac{r-1}{Nr-r+1}
    \geq \frac{r-1}{rN^3}.
  \end{align}
  Similarly,
  \begin{align}
    \frac{1}{N}\frac{r}{w_v(S)} - \frac{1}{w(S)}\frac{1}{\deg(v)}
    &\geq \frac{r}{rN\deg(v)} - \frac{1}{(N-1+r)\deg(v)}
    = \frac{1}{\deg(v)}\frac{1}{N}\frac{r-1}{N+r-1}
    \geq \frac{r-1}{rN^3}.
  \end{align}
  Thus $\psi^{\delta=1/2}_{u,v}(S) \geq \frac{r-1}{rN^3}$, and the absorption time bound follows from \Cref{theorem: absorption time advantage}.
\end{proof}

\begin{lemma}[Coupling with fitnesses $1$ and $r$ at fixed $\delta$]
  \label{lem:coupling-r}
  Fix an undirected unweighted graph $G=(V,E)$ on $N$ vertices and let $\delta\in[0,1]$.
  There exists a coupling of two mixed $\delta$-updating chains,
  $S_t^{(1)}$ with fitness $r_1=1$ and $S_t^{(2)}$ with fitness $r_2\geq r_1$,
  starting from the same $S_0\subseteq V$,
  such that $S_t^{(2)} \supseteq S_{t}^{(1)}$ for all $t\ge 0$.
\end{lemma}
\begin{proof}
  The claim holds at $t=0$.
  Assume inductively that $S_\ell^{(2)} \supseteq S_\ell^{(1)}$ for all $\ell \leq t$.
  At step $t+1$, draw $D\sim\mathrm{Bernoulli}(\delta)$.

  In both cases, let $\mu_1$ and $\mu_2$ be the laws of the reproducing vertex in chains $1$ and $2$, and couple
  the two chains through a \emph{maximal} coupling of $\mu_1$ and $\mu_2$.
  We verify in each case that
  \begin{equation}
    \label{eq:coupling-domination}
    \mu_2(x)\geq\mu_1(x)\ \text{ for every }x\in S_t^{(2)},
    \qquad
    \mu_2(x)\leq\mu_1(x)\ \text{ for every }x\notin S_t^{(2)} .
  \end{equation}
  Under~\eqref{eq:coupling-domination}, the two chains draw the same vertex $x$ with probability
  $\sum_x\min\{\mu_1(x),\mu_2(x)\}$. Otherwise chain~$1$ draws from the excess of $\mu_1$ over $\mu_2$, which is
  supported outside $S_t^{(2)}$, and chain~$2$ from the excess of $\mu_2$ over $\mu_1$, supported inside $S_t^{(2)}$.

  \textbf{Bd case ($D=0$).}
  Here $\mu_i(x)=r_i/w(S_t^{(i)})$ for $x\in S_t^{(i)}$ and $\mu_i(x)=1/w(S_t^{(i)})$ otherwise, where
  $w(S)=r_i|S|+N-|S|$.
  Since $r_1=1$ we have $\mu_1\equiv1/N$, and~\eqref{eq:coupling-domination} reduces to
  $r_2/w(S_t^{(2)})\geq1/N\geq1/w(S_t^{(2)})$, which holds because
  $r_2N-w(S_t^{(2)})=(r_2-1)(N-|S_t^{(2)}|)\geq0$ and $w(S_t^{(2)})\geq N$.
  When the two chains draw the same vertex $x$, let them pick the same target neighbor $y$.
  If $x\in S_t^{(1)}$ then $y$ becomes a mutant in both chains. If $x\in S_t^{(2)}\setminus S_t^{(1)}$ then $y$ becomes
  wild-type in chain~$1$ and a mutant in chain~$2$. If $x\notin S_t^{(2)}$ then $y$ becomes wild-type in both.
  In each case $S_{t+1}^{(2)}\supseteq S_{t+1}^{(1)}$.
  When they draw different vertices, chain~$1$'s parent is wild-type and chain~$2$'s is a mutant, so chain~$1$ can
  only lose a mutant while chain~$2$ can only gain one. Pick the targets independently and again
  $S_{t+1}^{(2)}\supseteq S_{t+1}^{(1)}$.

  \textbf{dB case ($D=1$).}
  Draw a common death vertex $v$ uniformly from $V$. This law is the same in both chains.
  Now $\mu_i$ is the law of the reproducing neighbor of $v$, namely $\mu_i(x)=r_i/w_v(S_t^{(i)})$ for
  $x\in S_t^{(i)}$ adjacent to $v$ and $\mu_i(x)=1/w_v(S_t^{(i)})$ for the remaining neighbors.
  Since $r_1=1$ we have $\mu_1\equiv1/\deg(v)$, and~\eqref{eq:coupling-domination} follows from
  $r_2\deg(v)-w_v(S_t^{(2)})=(r_2-1)\big(\deg(v)-\deg_{S_t^{(2)}}(v)\big)\geq0$ and $w_v(S_t^{(2)})\geq\deg(v)$.
  Since $v$ is common to both chains, the case analysis of the Bd case applies verbatim with $y=v$, and gives
  $S_{t+1}^{(2)}\supseteq S_{t+1}^{(1)}$.
\end{proof}

\begin{corollary}
  \label{corollary:lower-bound-fp-delta-1/2}
  For an undirected unweighted graph $G=(V,E)$ with $N$ vertices and $S_0\subseteq V$ with $r>1$,
  we have $\mathsf{fp}_{r}^{\delta=1/2}(G, S_0)\geq |S_0|/N$.
\end{corollary}
\begin{proof}
  By \Cref{lem:coupling-r}, for any $t$, $S_t^{(2)}\supseteq S_t^{(1)}$.
  Since $\mathsf{fp}_{r=1}^{\delta=1/2}(G,S_0) = |S_0|/N$ (by \Cref{thm:neutral-evolution} together with the additivity result~\Cref{claim:fp-additive}),
  and fixation of chain $1$ implies fixation of chain $2$, the claim follows.
\end{proof}

\begin{proof}[Proof of \Cref{thm:fixation-time-selection-half}]
  We know $\mathsf{FT}_{r}^{\delta}(G,S_0)\leq \mathsf{AT}_r^{\delta}(G,S_0) / \mathsf{fp}_r^\delta(G,S_0)$.
  By \Cref{thm:abs-time-advantageous-half}, $\mathsf{AT}_r^{\delta=1/2}(G,S_0)\leq \frac{r}{r-1}\cdot N^4$.
  By \Cref{corollary:lower-bound-fp-delta-1/2}, $\mathsf{fp}_{r}^{\delta=1/2}(G,S_0)\geq |S_0|/N\geq 1/N$.
  Thus $\mathsf{FT}_r^{\delta=1/2}(G,\{u\})\leq \frac{r}{r-1}\cdot N^4\cdot N = \frac{r}{r-1}\cdot N^5$.
\end{proof}

\section{Proof of \Cref{thm:cycles}}
\label{sec:cycle}

Throughout this section $r>0$ and $\delta\in[0,1]$ are arbitrary.
We consider the fixation probability on a cycle $C_N$ with $N$ vertices starting with one mutant.
Since the current number of mutants completely determines the state (mutants are all adjacent),
we only need to compute the ratio $\gamma_k$ of the probability of decreasing over the probability of increasing
when there are $k$ mutants.
From this, the standard formula (see~\cite{nowak2006evolutionary}) gives
\begin{equation}
  \mathsf{fp}_{r}^{\delta}(C_N) = \frac{1}{1 + \sum_{j=1}^{N-1}\prod_{k=1}^j \gamma_k}.
\end{equation}
Let $F_k := rk + (N-k)$ be the total fitness of the population with $k$ mutants.
The probability of increasing the mutant count with $k<N-1$ mutants is
\begin{equation}
  p_k^{\uparrow} = (1-\delta) r/F_k + 2r\delta/((1+r)N).
\end{equation}
When $k=N-1$,
\begin{equation}
  p_{N-1}^{\uparrow} = (1-\delta) r/F_{N-1} + \delta/N.
\end{equation}
The probability of decreasing with $k>1$ mutants is
\begin{equation}
  p_k^{\downarrow} = (1-\delta)/F_k + 2\delta/((1+r)N).
\end{equation}
When $k=1$,
\begin{equation}
  p_1^{\downarrow} = (1-\delta)/F_1 + \delta/N.
\end{equation}
Thus
\begin{equation}
  \gamma_k = p_k^{\downarrow}/p_k^{\uparrow} =
  \begin{cases}
    \displaystyle\frac{(1-\delta)/F_1 + \delta/N}{(1-\delta) r/F_1 + 2r\delta/((1+r)N)}&\text{if }k=1,\\[10pt]
    \displaystyle\frac{(1-\delta)/F_{N-1} + 2\delta/((1+r)N)}{(1-\delta) r/F_{N-1} + \delta/N}&\text{if }k=N-1,\\[10pt]
    \displaystyle\frac{(1-\delta)/F_k + 2\delta/((1+r)N)}{(1-\delta) r/F_k + 2r\delta/((1+r)N)}&\text{otherwise.}
  \end{cases}
\end{equation}

\section{Fixation probabilities on stars (Proofs of \Cref{thm:stars,thm:stars-selection})}
\label{sec:star-selection}
\newcommand{\Ps}[1]{P_{#1}^{\star}}
\newcommand{\Pe}[1]{P_{#1}^{\varnothing}}
\newcommand{\Fstar}[1]{F_{#1}^{\star}}
\newcommand{\Femp}[1]{F_{#1}^{\varnothing}}
\newcommand{\Gi}{G_i}

We derive an explicit formula for a star graph for general $r>0$ and $\delta\in[0,1]$.
We consider a star graph $S_N$ with a center vertex labelled $c$ and $n=N-1$ leaves labelled $\ell_1,\ldots,\ell_n$.
A state in our Markov chain is $(i,\sigma)$
where $i\in\{0,\dots,n\}$ counts mutant leaves and
$\sigma\in\{\star,\varnothing\}$ records whether the center is mutant ($\star$) or wild-type ($\varnothing$).
We write
\begin{equation}
  \Ps{i}:=\mathbb P\{\text{fixation}\mid(i,\star)\},
  \qquad
  \Pe{i}:=\mathbb P\{\text{fixation}\mid(i,\varnothing)\}.
\end{equation}
The boundary conditions are $\Pe{0}=0$ and $\Ps{n}=1$.

Let $d_i:=n-i$ be the number of wild-type leaves.
The three fitness totals are
\begin{equation}
  \Fstar{i}=ri+d_i+r,\qquad
  \Femp{i}=ri+d_i+1,\qquad
  \Gi=ri+d_i.
\end{equation}

Following the analysis in~\cite{broom2008analysis}, at a center mutant state the process satisfies
\begin{equation}
  (1-C_i)\Ps{i}=A_i\Ps{i+1}+B_i\Pe{i},
\end{equation}
where
\begin{align}
  A_i &= (1-\delta)\frac{rd_i}{n\Fstar{i}}+\delta\frac{d_i}{N},\\
  B_i &= (1-\delta)\frac{d_i}{\Fstar{i}}+\delta\frac{d_i}{N\Gi},\\
  C_i &= (1-\delta)\Bigl(\frac{ri}{\Fstar{i}}+\frac{ri}{n\Fstar{i}}\Bigr)
       +\delta\Bigl(\frac{i}{N}+\frac{ri}{N\Gi}\Bigr).
\end{align}
Solving gives $\Ps{i}=\alpha_i\Ps{i+1}+\beta_i\Pe{i}$ with $\alpha_i=A_i/(1-C_i)$ and $\beta_i=B_i/(1-C_i)$.

At a center wild-type state,
\begin{equation}
  (1-b_i)\Pe{i}=a_i\Ps{i}+c_i\Pe{i-1},
\end{equation}
where
\begin{align}
  a_i &= (1-\delta)\frac{ri}{\Femp{i}}+\delta\frac{ri}{N\Gi},\\
  c_i &= (1-\delta)\frac{i}{n\Femp{i}}+\delta\frac{i}{N},\\
  b_i &= 1-a_i-c_i.
\end{align}
Solving gives $\Pe{i}=p_i\Ps{i}+q_i\Pe{i-1}$ with $p_i=a_i/(1-b_i)$ and $q_i=c_i/(1-b_i)$.

These two recurrences combine into the matrix form
\begin{equation}
  \begin{pmatrix}\Ps{i+1}\\\Pe{i}\end{pmatrix}
  = M_i\begin{pmatrix}\Ps{i}\\\Pe{i-1}\end{pmatrix},
  \quad
  M_i = \begin{pmatrix}\frac{1-\beta_ip_i}{\alpha_i}&-\frac{\beta_iq_i}{\alpha_i}\\p_i&q_i\end{pmatrix}.
\end{equation}
Setting $A^{(i)}:=M_iM_{i-1}\cdots M_1$ for $i=1,\dots,n-1$
(the order matters: $M_1$ acts first, and the $M_j$ do not commute unless $r=1$),
the boundary conditions give
\begin{equation}
  \Ps{1} = 1/A^{(n-1)}_{1,1},
\end{equation}
and then
\begin{equation}
  \mathsf{fp}_{r}^\delta(S_N, \{c\}) = \Ps{0} = \alpha_0\Ps{1},
  \qquad
  \mathsf{fp}_{r}^\delta(S_N, \{\ell_i\}) = \Pe{1} = p_1\Ps{1}.
\end{equation}
When $r=1$, the coefficients simplify and we recover the neutral evolution formulas of \Cref{thm:stars}, as we now show.

\subsection{Neutral specialization ($r=1$): proof of \Cref{thm:stars}}
\label{sec:star-neutral}
\begin{proof}
Let $d_i=n-i$ denote the number of wild-type leaves.
At $r=1$,
\begin{equation}
\Fstar{i}=\Femp{i}=n+1,\qquad \Gi=n,
\end{equation}
so the coefficients $A_i,B_i,C_i,a_i,c_i,b_i$ from the general derivation above simplify to
\begin{align}
A_i&=(1-\delta)\frac{d_i}{n(n+1)}+\delta\frac{d_i}{n+1}
=\frac{n-i}{n+1}\Big(\frac{(1-\delta)}{n}+\delta\Big),\\
B_i&=(1-\delta)\frac{d_i}{n+1}+\delta\frac{d_i}{n(n+1)}
=\frac{n-i}{n+1}\Big((1-\delta)+\frac{\delta}{n}\Big),\\
C_i&=\frac{i}{n},\\
a_i&=(1-\delta)\frac{i}{n+1}+\delta\frac{i}{n(n+1)}
=\frac{i}{n+1}\Big((1-\delta)+\frac{\delta}{n}\Big),\\
c_i&=(1-\delta)\frac{i}{n(n+1)}+\delta\frac{i}{n+1}
=\frac{i}{n+1}\Big(\frac{(1-\delta)}{n}+\delta\Big),\\
b_i&=1-a_i-c_i=1-\frac{i}{n}.
\end{align}
Normalizing gives
\begin{equation}
\alpha_i=\frac{A_i}{1-C_i},\quad
\beta_i=\frac{B_i}{1-C_i},\quad
p_i=\frac{a_i}{1-b_i},\quad
q_i=\frac{c_i}{1-b_i}.
\end{equation}
Since $1-C_i=(n-i)/n$ and $1-b_i=i/n$, the factors $(n-i)$ and $i$ cancel, and all four quantities become
\begin{equation}
\alpha=\frac{n\delta+(1-\delta)}{n+1},\qquad
\beta=\frac{n(1-\delta)+\delta}{n+1},\qquad
p=\beta,\qquad q=\alpha,\qquad \alpha+\beta=1.
\end{equation}

With these constants, the two recurrences read
\begin{equation}
\Ps{i}=\alpha\Ps{i+1}+\beta\Pe{i},
\qquad
\Pe{i}=\beta\Ps{i}+\alpha\Pe{i-1}.
\end{equation}
This can be written in $2\times2$ form as
\begin{equation}
  \begin{pmatrix}\Ps{i+1}\\ \Pe{i}\end{pmatrix}
  = M\begin{pmatrix}\Ps{i}\\ \Pe{i-1}\end{pmatrix},
  \qquad
  M=\begin{pmatrix}
  \frac{1-\beta p}{\alpha}&-\frac{\beta q}{\alpha}\\[2pt]
  p&q
  \end{pmatrix}
  =\begin{pmatrix}
  1+\beta&-\beta\\
  \beta&\alpha
  \end{pmatrix}.
\end{equation}
We will raise this matrix to a power.
Note that
\begin{equation}
\mathrm{tr}(M)=(1+\beta)+\alpha=2,\qquad \det(M)=(1+\beta)\alpha-(-\beta)\beta=(1+\beta)(1-\beta)+\beta^2=1,
\end{equation}
so $M$ has eigenvalue $1$ with algebraic multiplicity $2$.
Indeed,
\begin{equation}
M-I=\begin{pmatrix}\beta&-\beta\\ \beta&-\beta\end{pmatrix}
=\beta \begin{pmatrix}1&-1\\ 1&-1\end{pmatrix}{=:Q},
\qquad
Q^2=0,
\end{equation}
hence $(M-I)^2=0$.
Therefore by the binomial theorem, for every $t\in\mathbb{Z}_{\ge 0}$,
\begin{equation}\label{eq:power}
M^t=(Q+I)^t=I+tQ
=\begin{pmatrix}
1+t\beta&-t\beta\\
t\beta&1-t\beta
\end{pmatrix}.
\end{equation}

Now apply the boundary conditions $\Pe{0}=0$ and $\Ps{n}=1$.
Iterating the relation
$(\Ps{i+1},\Pe{i})^\top=M(\Ps{i},\Pe{i-1})^\top$
gives
\begin{equation}
\begin{pmatrix}\Ps{n}\\ \Pe{n-1}\end{pmatrix}
=M^{n-1}\begin{pmatrix}\Ps{1}\\ \Pe{0}\end{pmatrix}
=M^{n-1}\begin{pmatrix}\Ps{1}\\ 0\end{pmatrix}.
\end{equation}
Taking the first coordinate and using \eqref{eq:power},
\begin{equation}
1=\Ps{n}=(M^{n-1})_{11}\Ps{1}=\big(1+(n-1)\beta\big)\Ps{1},
\end{equation}
so
\begin{equation}\label{eq:Ps1}
\Ps{1}=\frac{1}{1+(n-1)\beta}.
\end{equation}
The identities $\Ps{0}=\alpha\Ps{1}$ and $\Pe{1}=p\Ps{1}=\beta\Ps{1}$ then give
\begin{equation}
\Ps{0}=\frac{\alpha}{\alpha+n\beta},\qquad
\Pe{1}=\frac{\beta}{\alpha+n\beta}.
\end{equation}
It is convenient to parametrize by $z:=\alpha/\beta$ (well-defined since $\beta>0$ for $n\ge1$).
Then
\begin{equation}
\Ps{0}=\frac{z}{n+z},\qquad \Pe{1}=\frac{1}{n+z}.
\end{equation}
A short calculation with the explicit $\alpha,\beta$ yields
\begin{equation}
z=\frac{\alpha}{\beta}=\frac{n\delta+(1-\delta)}{(1-\delta)(n-1)+1},
\end{equation}
and the general formulas
\begin{equation}
\Pe{i}=\frac{i}{n+z},\qquad
\Ps{i}=\frac{i+z}{n+z}
\end{equation}
follow by a single step induction using the two recurrences (they also match the boundary values $\Pe{0}=0,\Ps{n}=1$).
Replacing $n=N-1$ gives the formulas in $N$.
\end{proof}

\section{Proofs of \Cref{thm:fp-lower-bound-selection,thm:fixation-time-selection}}
\label{sec:strong-selection}

\begin{definition}[Almost regular]
  An unweighted undirected graph is $\lambda$-almost regular if its maximum degree is at most $\lambda$ times its minimum degree.
\end{definition}

\begin{theorem}
  \label{thm:abs-time-advantage}
  For $\lambda\geq 1$, a $\lambda$-almost regular graph $G=(V,E)$ on $N$ vertices with $r > \lambda^2$
  (so that in particular $r>1$),
  the mixed $\delta$-updating process has the following properties for all $\emptyset \subsetneq S\subsetneq V$:
  \begin{enumerate}
    \item\label{item:abs-time-advantage-abs-time} $\mathsf{AT}_r^{\delta}(G, S)\leq 4\cdot\frac{r}{r-1}\cdot N^5$, provided $N\geq 4$; and
    \item\label{item:abs-time-advantage-fp} $\mathsf{fp}_r^\delta(G, S)\geq|S|/N^2$.
  \end{enumerate}
\end{theorem}
\begin{proof}
  For a potential $\phi$ and a crossing edge $(u,v)$, that is, $(u,v)\in E$ with $u\in S$ and $v\notin S$, abbreviate
  $\Delta^{+}=\phi(S\cup\{v\})-\phi(S)$ and $\Delta^{-}=\phi(S)-\phi(S\setminus\{u\})$, so that
  \begin{equation}
    \label{eq:psi-abbrev}
    \psi_{u,v}^{\mathrm{Bd}}(S)=\frac{1}{w(S)}\left(\frac{r\Delta^{+}}{\deg(u)}-\frac{\Delta^{-}}{\deg(v)}\right),
    \qquad
    \psi_{u,v}^{\mathrm{dB}}(S)=\frac{1}{N}\left(\frac{r\Delta^{+}}{w_v(S)}-\frac{\Delta^{-}}{w_u(S)}\right),
  \end{equation}
  and the expected one step change satisfies
  \begin{equation}
    \label{eq:drift-decomposition}
    \mathbb{E}\big[\phi(Y_{i+1}) - \phi(Y_i) \mid Y_i = S\big]
    = \sum_{(u,v)\ \mathrm{crossing}} \psi_{u,v}^{\delta}(S).
  \end{equation}
  Since the claim is for a fixed $\delta$, we may use a different potential in each of the two regimes
  $\delta\leq 1/2$ and $\delta\geq 1/2$.

  \medskip
  \noindent\textbf{Regime 1 ($\delta \leq 1/2$).}
  Take $\phi_1(S)=|S|$, so that $\Delta^{+}=\Delta^{-}=1$.
  If $\delta=1/2$, then $\psi_{u,v}^{\delta=1/2}(S) \geq \frac{r-1}{rN^3}$ by \Cref{thm:abs-time-advantageous-half}.
  If $\delta<1/2$, write $\psi_{u,v}^{\delta}(S) = (1-2\delta)\psi_{u,v}^{\mathrm{Bd}}(S) + 2\delta\psi_{u,v}^{\delta=1/2}(S)$,
  which is a convex combination since $1-2\delta\geq0$, and note
  \begin{equation}
    \psi_{u,v}^{\mathrm{Bd}}(S)
    = \frac{1}{w(S)}\left(\frac{r}{\deg(u)} - \frac{1}{\deg(v)}\right)
    \geq \frac{1}{rN}\cdot\frac{r-\deg(u)/\deg(v)}{\deg(u)}
    \geq \frac{r-\lambda}{rN^2}.
  \end{equation}
  Since $r>\lambda^2$, we have $r-1\geq(\lambda-1)(\lambda+1)$, so $r-\lambda\geq(r-1)/N$ for $N\geq 2$,
  giving $\psi_{u,v}^{\mathrm{Bd}}(S)\geq\frac{r-1}{rN^3}$.
  Either way $\psi_{u,v}^{\delta}(S)\geq\frac{r-1}{rN^3}$ for every crossing edge, and since $S\notin\{\emptyset,V\}$
  there is at least one, so by~\eqref{eq:drift-decomposition} the drift of $\phi_1$ is at least $\frac{r-1}{rN^3}$.
  \Cref{theorem: absorption time advantage} with $k_1=\max_S\phi_1(S)=N$ and $k_2=\frac{r-1}{rN^3}$ gives
  $\mathbb E[\tau]\leq\frac{r}{r-1}N^4$, which is at most $4\frac{r}{r-1}N^5$.

  \medskip
  \noindent\textbf{Regime 2 ($\delta \geq 1/2$).}
  Take $\phi_2(S)=\sum_{u\in S}\deg(u)$, so that $\Delta^{+}=\deg(v)$ and $\Delta^{-}=\deg(u)$.
  Note $\phi_2(\emptyset)=0$ and $\phi_2(S)=\phi_2(V)$ only for $S=V$.

  \emph{Step 1 (the drift is never negative).}
  Every fitness is at least $1$ and at most $r$, so $\deg(u)\leq w_u(S)$ and $w_v(S)\leq r\deg(v)$, whence
  \begin{equation}
    \label{eq:db-nonneg}
    \frac{r\deg(v)}{w_v(S)}\geq 1\geq\frac{\deg(u)}{w_u(S)}
    \qquad\text{and so}\qquad
    \psi_{u,v}^{\mathrm{dB}}(S)\geq 0,
  \end{equation}
  with equality in~\eqref{eq:db-nonneg} exactly when $w_v(S)=r\deg(v)$ and $w_u(S)=\deg(u)$, that is, when every
  neighbor of $v$ lies in $S$ and every neighbor of $u$ lies outside $S$.
  Call such a crossing edge \emph{stalled}.
  Also $\deg(v)/\deg(u)\geq 1/\lambda$ and $\deg(u)/\deg(v)\leq\lambda$, so
  \begin{equation}
    \psi_{u,v}^{\mathrm{Bd}}(S)
    =\frac{1}{w(S)}\left(\frac{r\deg(v)}{\deg(u)}-\frac{\deg(u)}{\deg(v)}\right)
    \geq\frac{1}{w(S)}\left(\frac{r}{\lambda}-\lambda\right)
    =\frac{r-\lambda^2}{\lambda\, w(S)}\geq 0 .
  \end{equation}
  Hence $\psi_{u,v}^{\delta}(S)\geq0$ for every crossing edge and every $\delta\in[0,1]$, and therefore the drift of $\phi_2$ is
  always non-negative.

  \emph{Step 2 (non-stalled edges have a uniform positive bound).}
  Suppose the crossing edge $(u,v)$ is non-stalled.
  If some neighbor of $v$ lies outside $S$, then $w_v(S)\leq \deg(v)+(r-1)(\deg(v)-1)=r\deg(v)-(r-1)$, so
  $\frac{r\deg(v)}{w_v(S)}\geq 1+\frac{r-1}{r\deg(v)}\geq 1+\frac{r-1}{rN}$, while $\frac{\deg(u)}{w_u(S)}\leq1$.
  Otherwise some neighbor of $u$ lies in $S$, so $w_u(S)\geq\deg(u)+(r-1)$ and
  $\frac{\deg(u)}{w_u(S)}\leq 1-\frac{r-1}{\deg(u)+r-1}\leq 1-\frac{r-1}{rN}$
  (using $\deg(u)+r-1\leq N-1+r\leq rN$), while $\frac{r\deg(v)}{w_v(S)}\geq1$.
  In both cases
  \begin{equation}
    \label{eq:db-nonstalled}
    \psi_{u,v}^{\mathrm{dB}}(S)\geq\frac{1}{N}\cdot\frac{r-1}{rN}=\frac{r-1}{rN^2},
    \qquad\text{hence}\qquad
    \psi_{u,v}^{\delta}(S)\geq\delta\,\frac{r-1}{rN^2}\geq\frac{r-1}{2rN^2}
  \end{equation}
  for $\delta\geq1/2$, using Step 1 for the Bd part.

  \emph{Step 3 (all crossing edges stalled forces a bipartition).}
  Suppose every crossing edge of $S$ is stalled.
  If $u\in S$ has a neighbor outside $S$ then all of $u$'s neighbors lie outside $S$. If $v\notin S$ has a neighbor
  in $S$ then all of $v$'s neighbors lie in $S$.
  So the set of vertices incident to a crossing edge is closed under taking neighbors, and as $G$ is connected it
  is all of $V$.
  Therefore every vertex of $S$ has all its neighbors in $V\setminus S$ and conversely, so $G$ is bipartite with parts
  $S$ and $V\setminus S$.
  A connected bipartite graph has only one bipartition, so no set obtained from $S$ by adding or removing a single
  vertex has all of its crossing edges stalled, unless that set is $\emptyset$ or $V$.

  \emph{Step 4 (drift over two steps).}
  Let $Z_i:=Y_{2i}$ and fix $S$ with $\emptyset\subsetneq S\subsetneq V$.
  By Step 1 every one step drift is non-negative, so
  $\mathbb E[\phi_2(Y_{2})-\phi_2(Y_0)\mid Y_0=S]
   = \mathbb E[\phi_2(Y_1)-\phi_2(Y_0)\mid Y_0=S] + \mathbb E\big[\mathbb E[\phi_2(Y_2)-\phi_2(Y_1)\mid Y_1]\big]$
  is bounded below by either summand.
  If $S$ has a non-stalled crossing edge, the first summand is at least $\frac{r-1}{2rN^2}$ by~\eqref{eq:db-nonstalled}.
  Otherwise, by Step 3, $S$ is one side of the bipartition of $G$.
  With probability $\delta\geq1/2$ the step is a dB step, in which case a uniformly chosen vertex $w$ changes type
  and $Y_1$ is $S\setminus\{w\}$ or $S\cup\{w\}$. This is a proper non-empty subset of $V$ unless $w$ is the unique
  element of $S$ or the unique element of $V\setminus S$, which happens with probability at most $2/N$.
  By Step 3 any such $Y_1$ has a non-stalled crossing edge, so its own drift is at least $\frac{r-1}{2rN^2}$.
  Hence in both cases
  \begin{equation}
    \mathbb E[\phi_2(Y_{2})-\phi_2(Y_0)\mid Y_0=S]\ \geq\ \frac12\left(1-\frac2N\right)\frac{r-1}{2rN^2}
    \ =\ \frac{(N-2)(r-1)}{4rN^3}\ =:\ k_2 .
  \end{equation}

  \emph{Step 5.}
  Apply \Cref{theorem: absorption time advantage} to $(Z_i)$ with $\phi_2$, $k_1=\phi_2(V)=\sum_{v\in V}\deg(v)\leq N^2$
  and $k_2$ as above, giving $\mathbb E[\tau_Z]\leq k_1/k_2\leq\frac{4rN^5}{(N-2)(r-1)}$.
  Since $Y_{2\tau_Z}$ is absorbed we have $\tau\leq2\tau_Z$, so
  $\mathbb E[\tau]\leq\frac{8rN^5}{(N-2)(r-1)}\leq 4\cdot\frac{r}{r-1}\cdot N^5$ whenever $N\geq4$.
  This proves item \ref{item:abs-time-advantage-abs-time}.

  \medskip
  \noindent\textbf{Item \ref{item:abs-time-advantage-fp}.}
  In either regime the potential $\phi\in\{\phi_1,\phi_2\}$ is non-negative with $\phi(\emptyset)=0$, its drift is
  non-negative by the above, and $\mathbb E[\tau]<\infty$ by \Cref{lem:tau-finite}.
  Apply \Cref{theorem: fixation prob selection} with $k_1=\phi(S)$ and $k_2=\phi(V)$.
  For $\delta\leq1/2$ this gives $\mathsf{fp}_r^\delta(G,S)\geq|S|/N\geq|S|/N^2$.
  For $\delta\geq1/2$ it gives
  \begin{equation}
    \mathsf{fp}_r^\delta(G,S)\geq\frac{\sum_{u\in S}\deg(u)}{\sum_{v\in V}\deg(v)}
    \geq\frac{|S|\min_u\deg(u)}{N\max_v\deg(v)}=\frac{|S|}{\lambda N}\geq\frac{|S|}{N^2},
  \end{equation}
  since $\lambda\leq N-1$.
\end{proof}

\begin{proof}[Proof of \Cref{thm:fp-lower-bound-selection}]
  Set $\lambda=D/d$ in \Cref{thm:abs-time-advantage}.
  Since $r>(D/d)^2=\lambda^2$ and the graph is $\lambda$-almost regular,
  item \ref{item:abs-time-advantage-fp} gives $\mathsf{fp}_r^\delta(G,\{u\})\geq 1/N^2$.
\end{proof}

\begin{proof}[Proof of \Cref{thm:fixation-time-selection}]
  Set $\lambda=D/d$ in \Cref{thm:abs-time-advantage} and let $N\geq4$.
  Since $r>\lambda^2$ and $\mathsf{fp}_r^\delta(G,S)\geq|S|/N^2\geq 1/N^2$ by \Cref{thm:fp-lower-bound-selection},
  we get $\mathsf{FT}_r^\delta(G,\{u\})\leq \mathsf{AT}_r^\delta(G,\{u\})/\mathsf{fp}_r^\delta(G,\{u\})\leq 4\cdot\frac{r}{r-1}\cdot N^5\cdot N^2=4\cdot\frac{r}{r-1}\cdot N^7$.
\end{proof}

\section{FPRAS Amplification}
\begin{theorem}[FPRAS powering lemma]
  \label{thm:fpras-amplification}
  If there is an FPRAS with confidence $1-c$ for some constant $c>0$, then there is an FPRAS with confidence $1-\alpha$ for $\alpha<c$.
\end{theorem}
\begin{proof}
  The idea is to run the FPRAS $b_c\cdot\lceil\log(1/\alpha)\rceil + 1$ times for some particular constant $b_c>0$.
  Then return the median of the runs.
  See Lemma 6.1 in~\cite{jerrum1986random} for the proof and how to choose $b_c$.
\end{proof}

\section{FPRAS for $\delta=1/2$}
\begin{theorem}[FPRAS for $\delta=1/2$]
\label{thm:estimation-algorithm-delta-1/2}
  There exists a randomized algorithm $\mathcal A$, running in time polynomial in $N$, $1/\varepsilon$, $\log(1/\alpha)$ and $r/(r-1)$,
  such that on input
  $r>1$,
  $\varepsilon>0$,
  $\alpha\in (0,1]$,
  an undirected unweighted graph $G=(V,E)$ with $N$ vertices,
  and $S_0\subseteq V$,
  outputs with probability at least $1-\alpha$ an estimate to the fixation probability
  within a relative error of $\varepsilon$.
  In other words,
  \begin{equation}
    \mathbb P\left\{\big|\mathcal A(G, S_0, r)-\mathsf{fp}_r^{\delta=1/2}(G, S_0)\big| > \varepsilon\cdot \mathsf{fp}_r^{\delta=1/2}(G, S_0) \right\} < \alpha.
  \end{equation}
  The probability is over sampling the randomness for $\mathcal A$.
\end{theorem}
\begin{proof}
 The algorithm that follows uses a similar technique to the proof of~\Cref{thm:estimation-algorithm} in~\Cref{sec:estimation-algo} below.
 Let $s=\lceil 2\ln(2/\alpha_1)/(\rho \varepsilon)^2\rceil$ where $\rho=|S_0|/N$ for $\alpha_1=1/8$.
 Let $T=\lceil \alpha_2^{-1}sN^4 r/(r-1)\rceil$ for $\alpha_2=1/8$.

 Consider $s$ independent copies of the mixed $1/2$-updating process that will run all the way to absorption.
 Let $\mathcal E_2$ be the event that at least one copy will not absorb within $T$ steps.
 First notice that by a union bound, Markov's inequality, and \Cref{thm:abs-time-advantageous-half},
 we have
 $\mathbb P\{\mathcal E_2\} < s\cdot \mathsf{AT}_r^{\delta=1/2}(G, S_0)/T < \alpha_2$.
 If $\mathcal E_2$ occurs, we output $0$ as our estimate for $\mathsf{fp}_{r}^{\delta=1/2}(G, S_0)$.
 Otherwise all copies will absorb.
 For each $i\in\{1,\dots,s\}$, let $Z_i$ be the indicator that the $i$-th copy will fixate, and set
 $\hat f := \frac1s \sum_{i=1}^s Z_i$.
 Then $Z_1,\dots,Z_s$ are i.i.d.\ Bernoulli random variables with
 $\mathbb E[Z_i] = \mathsf{fp}_r^{\delta=1/2}(G,S_0)$.
 Moreover, on the event $\neg\mathcal E_2$, all $s$ copies will absorb by time $T$,
 so in that case the algorithm observes all absorption outcomes and outputs $\hat f$.
 Hence the event \textsf{FAIL} that the algorithm's output deviates from $\mathsf{fp}_r^{\delta=1/2}(G,S_0)$ by more than a multiplicative factor $\varepsilon$ satisfies
 \begin{equation}
   \textsf{FAIL}\subseteq \left(\mathcal E_2 \cup \Big\{|\hat f - \mathsf{fp}_r^{\delta=1/2}(G,S_0)| > \varepsilon\cdot \mathsf{fp}_r^{\delta=1/2}(G,S_0)\Big\}\right).
 \end{equation}
 By a Chernoff bound and \Cref{corollary:lower-bound-fp-delta-1/2}, we obtain
 \begin{align}
   \mathbb P\Big\{|\hat f - \mathsf{fp}_r^{\delta=1/2}(G,S_0)| > \varepsilon \cdot \mathsf{fp}_r^{\delta=1/2}(G,S_0)\Big\}
   <
   2\exp(-s\cdot (\rho\varepsilon)^2/2)
   < \alpha_1.
 \end{align}
 Therefore by a union bound, the probability that our algorithm errs is at most
 \begin{equation}
    \mathbb P\{\mathsf{FAIL}\}
    <
    \mathbb P\{\mathcal E_2\} + \alpha_1
    < \alpha_1 + \alpha_2 =1/4.
 \end{equation}

 We can check if $\mathcal E_2$ occurs by simulating each process for $T$ time steps and checking if any process has not absorbed.
 The running time of this algorithm is polynomial in $s$ and $T$,
 which are both polynomial in $N$, $1/\varepsilon$ and $r/(r-1)$.
 Finally, we use~\Cref{thm:fpras-amplification} which contributes a multiplicative $O(\log(1/\alpha))$
 factor to our algorithm time.
 Thus the running time of the complete algorithm is polynomial in $N$, $1/\varepsilon$, $\log(1/\alpha)$ and $r/(r-1)$. 
\end{proof}

\section{Proof of \Cref{thm:estimation-algorithm}}
\label{sec:estimation-algo}

\begin{proof}[Proof of \Cref{thm:estimation-algorithm}]
The algorithm simulates the mixed $\delta$-updating process many times and records how many trajectories fixated.
Let $\rho=|S_0|/N^2$ (a lower bound on the fixation probability from \Cref{thm:fp-lower-bound-selection}).
Let $s=\lceil 2\ln(2/\alpha_1)/(\rho\varepsilon)^2\rceil$ where $\alpha_1=1/8$.
Let $T=\lceil 4\alpha_2^{-1}sN^5 r/(r-1)\rceil$ where $\alpha_2=1/8$
(the factor $4$ matches the constant in \Cref{thm:abs-time-advantage}, and affects neither the asymptotics nor the
claimed running time).

Consider $s$ independent copies of the mixed $\delta$-updating process that will run all the way to absorption.
Let $\mathcal E_2$ be the event that at least one copy will not absorb within $T$ steps.
By a union bound, Markov's inequality, and \Cref{thm:abs-time-advantage},
\begin{equation}
  \mathbb P\{\mathcal E_2\} < s\cdot \mathsf{AT}_r^\delta(G, S_0)/T < \alpha_2.
\end{equation}
If $\mathcal E_2$ occurs, we output $0$ as our estimate for $\mathsf{fp}_{r}^{\delta}(G, S_0)$.
Otherwise all copies absorb.
For each $i\in\{1,\dots,s\}$, let $Z_i$ be the indicator that the $i$-th copy fixates, and set
$\hat f := \frac{1}{s}\sum_{i=1}^s Z_i$.
Then $Z_1,\dots,Z_s$ are i.i.d.\ Bernoulli random variables with
$\mathbb E[Z_i] = \mathsf{fp}_r^{\delta}(G,S_0)$.
The event \textsf{FAIL} that the output deviates from $\mathsf{fp}_r^{\delta}(G,S_0)$ by more than $\varepsilon$ satisfies
\begin{equation}
  \textsf{FAIL}\subseteq \left(\mathcal E_2 \cup \Big\{|\hat f - \mathsf{fp}_r^{\delta}(G,S_0)| > \varepsilon\cdot \mathsf{fp}_r^{\delta}(G,S_0)\Big\}\right).
\end{equation}
By a Chernoff bound and \Cref{thm:fp-lower-bound-selection},
\begin{align}
  \mathbb P\Big\{|\hat f - \mathsf{fp}_r^{\delta}(G,S_0)| > \varepsilon\cdot\mathsf{fp}_r^{\delta}(G,S_0)\Big\}
  < 2\exp(-s\cdot(\rho\varepsilon)^2/2) < \alpha_1.
\end{align}
Therefore
\begin{equation}
  \mathbb P\{\mathsf{FAIL}\} < \mathbb P\{\mathcal E_2\} + \alpha_1 < \alpha_1 + \alpha_2 = 1/4.
\end{equation}
The running time is polynomial in $s$ and $T$, both polynomial in $N$, $1/\varepsilon$ and $r/(r-1)$.
Applying \Cref{thm:fpras-amplification} contributes a multiplicative $O(\log(1/\alpha))$ factor,
giving total runtime polynomial in $N$, $1/\varepsilon$, $\log(1/\alpha)$ and $r/(r-1)$.
\end{proof}

\end{document}